\documentclass[aps,reprint]{revtex4-1}
\usepackage{amsfonts}
\usepackage{amsmath}
\usepackage{amssymb}
\usepackage{graphicx}%
\usepackage{graphics}
\providecommand{\U}[1]{\protect\rule{.1in}{.1in}}

\begin{document}
\title{EIT-related phenomena and their mechanical analogs}
\author{J. A. Souza}
\email{jamesfisica@gmail.com}
\affiliation{Departamento de F\'{\i}sica, Qu\'{\i}mica e Matem\'{a}tica, Universidade
Federal de S\~{a}o Carlos, Sorocaba, SP 18052-780, Brazil}
\affiliation{Departamento de F\'{\i}sica, Universidade Federal de S\~{a}o Carlos, S\~{a}o
Carlos, SP 13565-905, Brazil}
\author{L. Cabral}
\affiliation{Departamento de F\'{\i}sica, Universidade Federal de S\~{a}o Carlos, S\~{a}o
Carlos, SP 13565-905, Brazil}
\author{R. R. Oliveira}
\affiliation{Departamento de F\'{\i}sica, Universidade Federal de S\~{a}o Carlos, S\~{a}o
Carlos, SP 13565-905, Brazil}
\author{C. J. Villas-Boas}
\affiliation{Departamento de F\'{\i}sica, Universidade Federal de S\~{a}o Carlos, S\~{a}o
Carlos, SP 13565-905, Brazil}
\date{\today}

\begin{abstract}
Systems of interacting classical harmonic oscillators have received considerable attention in the last years as analog models for describing electromagnetically induced transparency (EIT) and associated phenomena. We review these models and investigate their validity for a variety of physical systems using two- and three-coupled harmonic oscillators. From the simplest EIT-$\Lambda$ configuration and two-coupled single cavity modes we show that each atomic dipole-allowed transition and a single cavity mode can be represented by a damped harmonic oscillator. Thus, we have established a one-to-one correspondence between the classical and quantum dynamical variables. We show the limiting conditions and the equivalent for the EIT dark state in the mechanical system. This correspondence is extended to other systems that present EIT-related phenomena. Examples of such systems are two- and three-level (cavity EIT) atoms interacting with a single mode of an optical cavity, and four-level atoms in a inverted-Y and tripod configurations. The established equivalence between the mechanical and the cavity EIT systems, presented here for the first time, has been corroborated by experimental data. The analysis of the probe response of all these systems also brings to light a physical interpretation for the expectation value of the photon annihilation operator $\left\langle a\right\rangle$. We show it can be directly related to the electric susceptibility of systems, the composition of which includes a driven cavity field mode.

\end{abstract}
\maketitle

%\pacs{03.67.-a, 03.67.Mn, 05.30.Rt}

\section{Introduction}

Electromagnetically Induced Transparency (EIT) is a quantum interference phenomenon responsible for canceling the absorption of a weak probe laser by applying a strong electromagnetic control field in the same medium. In the last decades, much attention has been paid to study EIT and related phenomena leading to many different applications \cite{Harris1997, Marangos1998, Fleischhauer2005}. In its simplest configuration, two electromagnetic fields excite an ensemble of three-level atoms in $\Lambda$ configuration and the optical properties of the atomic medium are described by the first-order complex electric susceptibility $\chi_{e}^{(1)}$. Its real part Re$\left\{\chi_{e}^{(1)}\right\} $ is related to the index of refraction of the medium, featured by a region of anomalous dispersion leading to very small group velocities \cite{Hau1999, Scully1999, Yashchuk1999}. The zero absorption window is described by the imaginary part Im$\left\{\chi_{e}^{(1)}\right\} $, which allows applications ranging from high-resolution spectroscopy \cite{Marangos1998} to atomic clocks \cite{Vanier2005}.

Mechanical and electric analogies of EIT in a $\Lambda$ configuration and their characteristics in equivalent systems have been noted since Alzar et al. \cite{Nussenzveig2002} reproduced the phenomenology of EIT using two coupled harmonic oscillators and RLC circuits. They were inspired by Hammer and Prentiss \cite{Prentiss1988}, who modeled classically the stimulated resonance Raman effect with a set of three coupled classical pendulums. Due to the considerable practical usefulness provided by the classical results, many efforts have been made towards representing EIT-related phenomena in different atomic systems using classical models \cite{Tokman2002, Huang2010, Serna2011, Huang2013}. Its importance has recently grown up even more owing the number of reported classical systems that follow the same dynamics, such as metamaterials \cite{Prosvirnin2008, Soukoulis2009, Bettiol2009, Giessen2009,
Soukoulis2011, Chen2012}, cavity optomechanics \cite{Vahala2007, Kippenberg2008, Kippenberg2010, Painter2011}, multiple coupled photonic crystal cavities \cite{Wong2009}, acoustic structures \cite{Sheng2010}, coupled resonant systems \cite{Zhang2013}, and so on.

To date, no completely correspondence between the quantum and classical models which yields a direct comparison between the results has been realized. We establish in this work, a one-to-one correspondence between the classical and quantum dynamic variables using two classical coupled harmonic oscillators to model EIT in $\Lambda$ configuration. We also show the role of a cavity mode in the mechanical system to model EIT-like phenomena observed in two coupled cavity modes and in systems comprised by a single two-level atom interacting with a single mode of a resonator considering two configurations, the driven cavity field and the driven atom. The analysis of the probe response for the driven cavity cases reveal that $\left\langle a\right\rangle $ is directly related to the electric susceptibility of the atom-cavity or cavity-cavity systems.

The classical correspondence is also established for EIT-like observed in four-level atoms in the inverted-Y and tripod configurations, and for the cavity EIT (CEIT) system, considering three coupled classical harmonic oscillators. For the atomic tripod configuration we compare the classical analog obtained here with the analog published recently \cite{Huang2013}, showing the validity of both for different set of parameters. The analog for the CEIT system is presented for the first time and the result is compared with an experiment performed with $N\sim15$ atoms \cite{Muecke2010}. We show the validity and the limiting conditions to reproduce the quantum results using the classical models. This work can be considerably useful to provide a general mapping of EIT-like systems into a variety of classical systems.

\section{Classical analog of EIT in different physical systems using two-coupled harmonic oscillators}

Coupled harmonic oscillators are an intuitive model used as close analog for many phenomena, including the stimulated resonance Raman effect \cite{Prentiss1988}, electromagnetic induced transparency \cite{Nussenzveig2002, Tokman2002, Huang2010, Serna2011, Huang2013}, time dependent Josephson phenomena \cite{Zimmerman1971}, adiabatic and nonadiabatic processes \cite{Xiong1988, Romanenko2009}, level repulsion \cite{Brentano1994}, strongly interacting quantum systems \cite{Novotny2010}, one-half spin dynamics \cite{Glaser2003, Devaud2006}, coherent quantum states \cite{McKibben1977, Briggs2012, Eisfeld2012}, among others.

EIT and their classical analogs can be obtained when suitable conditions are prescribed. In what follows, we will briefly review some of the EIT-related systems and derive their linear electric susceptibilities from the density matrix formalism. Our focus is to show how the behavior of the electric susceptibility of each atomic system can be reproduced using coupled
oscillators, through the concept of mechanical susceptibility.

\subsection{The phenomenology of EIT reproduced in two-coupled harmonic oscillators}\label{SecEIT}

\begin{figure}%[!ht]
[ptbh]
\begin{center}
\includegraphics[width=8cm]%
{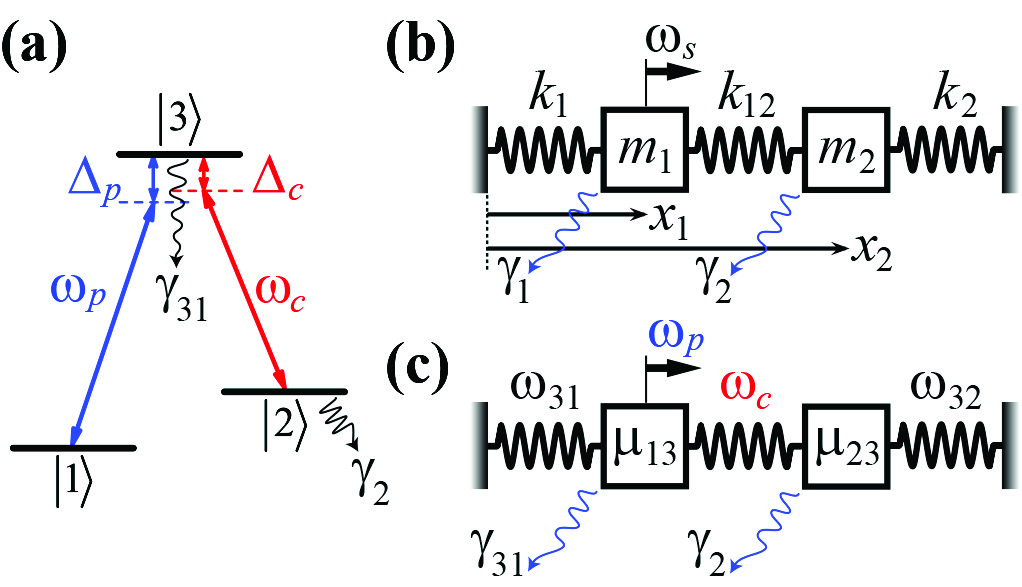}
\caption{(Color online). (a) Schematic energy level diagram of a three-level atom in $\Lambda$ configuration for EIT. It shows two classical electromagnetic fields, probe $(\omega_{p})$ and control $(\omega_{c})$, coupling the transitions $|1\rangle\leftrightarrow|3\rangle$ and $|2\rangle\leftrightarrow|3\rangle$, respectively, and their corresponding detunings. The decay rates are
represented by $\gamma_{31}=\Gamma_{31}+\Gamma_{32}+\gamma_{3}$ and $\gamma_{2}$. (b) Coupled damped harmonic oscillators used to reproduce the phenomenology observed in EIT, showing two masses $m_{1}$ and $m_{2}$ displaced from their equilibrium positions by the distances $x_{1}$ and $x_{2}$, respectively, attached to three springs with spring constants $k_{1}%
$, $k_{2}$ and $k_{12}$. A driving force of frequency $\omega_{s}$ acts on mass $m_{1}$ and the damping constant of the $j$th harmonic oscillator is represented by $\gamma_{j}$ ($j=1,2$). (c) Classical analog of EIT showing the equivalence of each parameter in the mechanical system. Each harmonic oscillator corresponds to a dipole-allowed transition with electronic dipole moment $\mu_{i3}$, ($i=1,2$).} \label{EsquemaEIT}
\end{center}
\end{figure}

The phenomenon of EIT occurs in three level atomic systems in $\Lambda$ configuration with two ground states, $|1\rangle$ and $|2\rangle$, and an excited state $|3\rangle$, interacting with two classical coherent fields, probe and control, of frequencies $\omega_{p}$ and $\omega_{c}$, respectively, as illustrated in Fig.\ref{EsquemaEIT}a. The atomic transition $|1\rangle \leftrightarrow|3\rangle$ (frequency $\omega_{31}$) is driven by the probe field with Rabi frequency $2\Omega_{p}$, and the transition $|2\rangle\leftrightarrow|3\rangle$ (frequency $\omega_{32}$) is coupled by the control field with Rabi frequency $2\Omega_{c}$ \cite{Multilevel}.

Introducing the electric dipole and rotating-wave approximations, the time-independent Hamiltonian which describes the atom-field interaction in a rotating frame is given by ($\hbar=1$) \cite{Fleischhauer2005}
\begin{equation}
H=(\Delta_{p}-\Delta_{c})\sigma_{22}+\Delta_{p}\sigma_{33}-(\Omega_{p}\sigma_{31}+\Omega_{c}\sigma_{32}+h.c.),\label{e0}
\end{equation}
where $\sigma_{ij}=\left\vert i\right\rangle \left\langle j\right\vert,i,j=1,2,3$ are the atomic raising and lowering operators ($i\neq j$), and atomic energy-level population operators ($i=j$). The detunings are given by $\Delta_{p}=\omega_{31}-\omega_{p}$, $\Delta_{c}=\omega_{32}-\omega_{c}$ and $h.c.$ stands for the Hermitian conjugate. The dynamics of the system is obtained by solving the master equation for the atomic density operator ($\rho$)
\begin{align}
\dot{\rho} &  =-i[H,\rho]+\sum\limits_{m=1,2}\Gamma_{3m}(2\sigma_{m3}\rho\sigma_{3m}-\sigma_{33}\rho-\rho\sigma_{33})\nonumber\\
&  +\sum\limits_{n=2,3}\gamma_{n}(2\sigma_{nn}\rho\sigma_{nn}-\sigma_{nn}\rho-\rho\sigma_{nn}),\label{e1}
\end{align}
where $\Gamma_{31}$, $\Gamma_{32}$ are the polarization decay rates of the excited level $|3\rangle$ to the levels $|1\rangle$ and $|2\rangle$, and $\gamma_{2}$, $\gamma_{3}$ the non-radiative atomic dephasing rates of states $|2\rangle$ and $|3\rangle$, respectively.

It is assumed that all $N$ atoms contained in a volume $V$ couple identically to the electromagnetic fields and that the medium is isotropic and homogenous. Considering that the atoms do not interact to each other and ignoring local-field effects, the optical response of the medium to the applied probe field $E(t)=E_{p}e^{-i\omega_{p}t}+c.c.$ can be obtained through the expectation value of the atomic polarizability
\begin{equation}
\mathbf{P}(t)=\chi_{e}^{(1)}\mathbf{E}(t), \label{e2}
\end{equation}
with $\chi_{e}^{(1)}$ denoting the linear electric susceptibility. The polarization can also be written in terms of the expectation value of the dipole moment operator $\mu$ per unit volume
\begin{equation}
\mathbf{P}(t)=-\frac{1}{V}\sum\limits_{i=1}^{N}\left\langle e\mathbf{r}_{i}(t)\right\rangle =\frac{N}{V}Tr(\mu\rho). \label{e3}
\end{equation}

In this way the linear response of the probe beam in the atomic sample can be directly related to the off-diagonal density matrix element $\rho_{31}$,
\begin{equation}
\chi_{e}^{(1)}(\omega_{p})=\frac{N\left\vert \mu_{13}\right\vert }{VE_{p}}\rho_{31}. \label{SFull}
\end{equation}

From eqs.\eqref{e0} and \eqref{e1} the full equations of motion for the density matrix are given by 
\begin{subequations}\label{EvolveEIT}
\begin{eqnarray}
\dot{\rho}_{31} &=& -i\left\{ \left(\Delta_{p} - i\gamma_{31}\right) \rho_{31} - \Omega_{p}\left(\rho_{11} - \rho_{33} \right)\right\} \nonumber\\
&+& i\Omega_{c}\rho_{21},\\
\nonumber\\
\dot{\rho}_{21} &=& -i\left\{ \left[ \left(\Delta_{p} - \Delta_{c}\right) - i\gamma_{2} \right] \rho_{21} + \Omega_{p} \rho_{23} \right\} \nonumber\\
&+& i\Omega_{c}\rho_{31},\\
\nonumber\\
\dot{\rho}_{23} &=& -i\left\{ \left[-\Delta_{c} - i\left( \gamma_{31} - \gamma_{2}\right)\right] \rho_{23} + \Omega_{p} \rho_{21} \right\} \nonumber\\
&+& i\Omega_{c}\left( \rho_{33} - \rho_{22} \right),
\end{eqnarray}
\end{subequations}
where $\gamma_{31} = \Gamma_{31} + \Gamma_{32} + \gamma_{3}$.

As described in detail by Fleischhauer et al. \cite{Fleischhauer2005}, EIT occurs when the population of the system is initially in the ground state $|1\rangle$. The state of zero absorption, referred to as the dark state, is usually attributed to the result of quantum interference between two indistinguishable paths. This state corresponds to $|1\rangle$ if the
conditions $\Omega_{p}<<\Omega_{c}$ and $\gamma_{2}<<\gamma_{31}$ are prescribed to yield $\rho_{11}\approx1$ and consequently $\rho_{22} \approx 0$. The state $|3\rangle$ is never populated ($\rho _{33} = 0$) in the dark state. Using these conditions in eqs.\eqref{EvolveEIT}, the steady-state solutions ($\dot{\rho}_{ij}=0$) for $\rho_{21}$ and $\rho_{31}$ can be determined through the equations
\begin{subequations}\label{EvolveEITB}
\begin{eqnarray}
\left(\Delta_{p} - i\gamma_{31}\right) \rho_{31} - \Omega_{c}\rho_{21} &=& \Omega_{p},\\
\nonumber\\
\left( \delta - i\gamma_{2} \right) \rho_{21} - \Omega_{c}\rho_{31} &=& 0,
\end{eqnarray}
\end{subequations}
yielding,
\begin{equation}
\rho_{31}(\omega_{p})=\frac{\Omega_{p}\left(\delta-i\gamma_{2}\right)}{\left(\Delta_{p}-i\gamma_{31}\right)\left(\delta-i\gamma_{2}\right)-\Omega_{c}^{2}}, \label{e5}%
\end{equation}
where we have introduced the two-photon detuning $\delta=\Delta_{p}-\Delta_{c}$.

Hereafter, the susceptibility stated in eq.\eqref{SFull} will be replaced by a reduced susceptibility that does not depend on the specific details of the physical system. Then, for EIT it reads,
\begin{equation}
\tilde{\chi}_{e}(\omega_{p})=\frac{VE_{p}}{N\left\vert \mu_{13}\right\vert}\chi_{e}^{(1)}(\omega_{p})=\rho_{31}(\omega_{p}). \label{e6}
\end{equation}

Thus, the main characteristics of EIT regarding absorption, gain and the control of the group velocity of light in a medium can be obtained from the imaginary and real parts of $\rho_{31}$.

Note that the essential features of EIT are derived using a semiclassical model, where it is assumed two classical fields interacting with an atomic ensemble with microscopic coherences treated quantum mechanically. Under the assumption of low atomic excitation ($\rho_{11}\approx1$), which is experimentally justified by choosing an appropriately low pump intensity, implying that $\Omega_{p}<<\Omega_{c}$, effects of atomic saturation are neglected. In this way, the expectation values of the atomic operators $\rho_{ij}=\left\langle \sigma_{ji}\right\rangle $ can be replaced by classical amplitudes.

The mechanical model used to demonstrate the classical analog of EIT consists of two coupled, damped harmonic oscillators with one of them driven by a harmonic force $F_{s}(t)=Fe^{-i\left(\omega_{s}t+\phi_{s}\right)}+c.c.$, for $\phi_{s}=0$ and frequency $\omega_{s}$ \cite{Nussenzveig2002}. It is considered two particles $1$ and $2$ with equal masses $m_{1}=m_{2}=m$ and three springs arranged as illustrated in Fig.\ref{EsquemaEIT} b. The two outside spring constants are $k_{1}$ and $k_{2}$. The third spring couples linearly the two particles and its constant spring is $k_{12}$. It is assumed that the whole system moves in only one dimension $x$ and the distances $x_{1}$ and $x_{2}$ measure the displacements of particles $1$ and $2$ from their respective equilibrium positions. The equations of motion for the two masses are
\begin{subequations}
\begin{eqnarray}
m\ddot{x}_{1} &=& -k_{1}x_{1} - \eta_{1}\dot{x}_{1} - k_{12}\left(x_{1} - x_{2}\right) + F_{s}(t)\\ \label{ClassA}
m\ddot{x}_{2} &=& -k_{2}x_{2} - \eta_{2}\dot{x}_{2} - k_{12}\left(x_{2} - x_{1}\right) \label{ClassB}
\end{eqnarray}
\end{subequations}
which are usually written as,
\begin{subequations}\label{classicA}
\begin{align}
\ddot{x}_{1} +\omega_{1}^{2}x_{1} + 2\gamma_{1}\dot{x}_{1} - \omega_{12}^{2}x_{2} &= \frac{F_{s}(t)}{m}\\
\ddot{x}_{2} +\omega_{2}^{2}x_{2} + 2\gamma_{2}\dot{x}_{2} - \omega_{12}^{2}x_{1} &= 0
\end{align}
\end{subequations}
where $\omega_{j}^{2} = \left(k_{j} + k_{12}\right)/m$, $\omega_{12}^{2} = k_{12}/m$ and the damping constant of the $j$th harmonic oscillator is $2\gamma_{j} = \eta_{j}/m$, $j = 1,2$. Assuming that the steady-state solution of equations above has the form $x_{j} = N_{j}e^{-i\omega_{s}t} + c.c.$ we find
\begin{subequations}\label{EvolveHO}
\begin{eqnarray}
\left(-\omega_{s}^2 + \omega_{1}^{2} - 2i\gamma_{1}\omega_{s}\right)N_{1} - \omega_{12}^{2}N_{2} &= \frac{F}{m},\\
\left(-\omega_{s}^2 + \omega_{2}^{2} - 2i\gamma_{2}\omega_{s}\right)N_{2} - \omega_{12}^{2}N_{1} &= 0,
\end{eqnarray}
\end{subequations}
where the complex conjugate solution ($c.c.$) was omitted for simplicity. Note that eqs.\eqref{EvolveHO} for $N_{1}$ and $N_{2}$ have the same struture as eqs.\eqref{EvolveEITB} for $\rho_{31}$ and $\rho_{21}$, respectively. 

Solving eqs.\eqref{EvolveHO} for the displacement of the driven oscillator $x_{1}\left(t\right)$ and considering $\omega_{s}$ near to the natural oscillation frequencies $\omega_{j}$ ($j=1,2$), so that $\omega_{j}^{2}-\omega_{s}^{2}\approx 2\omega_{j}(\omega_{j}-\omega_{s})$ and $\gamma_{j}\omega_{s} \approx \gamma_{j}\omega_{j}$, we have
\begin{equation}
x_{1}(t)\simeq\frac{F/\left(2m\omega_{1}\right) \left(\Delta_{2} - i\gamma_{2}\right)}{\left(\Delta_{1}-i\gamma_{1}\right) \left(\Delta_{2}-i\gamma_{2}\right)  -\Omega_{12}^{2}}e^{-i\omega_{s}t}+c.c., \label{e7}
\end{equation}
where we have defined the detunings $\Delta_{j}=\omega_{j}-\omega_{s}$ and the classical coupling rate between particles $1$ and $2$ as $2\Omega_{12}=\omega_{12}^{2}/\sqrt{\omega_{1}\omega_{2}}$, in analogy to the Rabi frequency of the control field ($2\Omega_{c}$). The quantity $F/m\omega_{1}=2\Omega_{s}C_{1}$ has dimension of frequency $(2\Omega_{s})$ times length $(C_{1})$. The first term makes the role of the Rabi frequency of the probe field ($2\Omega_{p}$). Then, eq.(\ref{e7}) can be reduced to the form
\begin{equation}
x_{1}(t)=C_{1}\rho_{co}e^{-i\omega_{s}t}+c.c., \label{e8}
\end{equation}
where the dimensionless complex amplitude $\rho_{co}$ is given by
\begin{equation}
\rho_{co}(\omega_{s})=\frac{\Omega_{s}\left(\Delta_{2}-i\gamma_{2}\right)}{\left(  \Delta_{1}-i\gamma_{1}\right)  \left(\Delta_{2}-i\gamma_{2}\right)  -\Omega_{12}^{2}}. \label{e9}
\end{equation}

An equation similar to \eqref{e8} can be derived for the atomic system by making $\left\vert \mathbf{r}_{i}(t)\right\vert =x(t)$ in eq.\eqref{e3} for $N=1$ and using eq.\eqref{e2}, eq.\eqref{SFull} and the expression for the applied probe field $E(t) = E_{p}e^{-i\omega_{p}t} + c.c.$, yielding,
\begin{equation}
x(t)=C_{2}\rho_{31}e^{-i\omega_{p}t}+c.c.=C_{2}\tilde{\chi}_{e}e^{-i\omega_{p}t}+c.c., \label{xc2}
\end{equation}
where $C_{2}=\left\vert \mu_{13}\right\vert /e$, similarly to $C_{1}$, bears dimension of length. By comparing eq.\eqref{e8} with the first equality of eq.\eqref{xc2} we find the analog $C_{1}\equiv C_{2}$, $\omega_{s}\equiv \omega_{p}$ and $\rho_{co}\equiv\rho_{31}$. The analog is obtained for the steady-state solution of both systems, EIT and coupled oscillators. In Appendix A we used the Hamiltonian formalism to show that the dynamics of the EIT system, given by $\dot{\rho}_{31}$ and $\dot{\rho}_{21}$, is also similar to the dynamics of the classical oscillators. This formalism is also advantageous to obtain a direct definition of the classical pumping rate $\Omega_{s}$ as a function of the parameters of the mechanical system, which is $\Omega_{s}=\sqrt{F^{2}/2m\omega_{1}}$ meaning that $C_{1}=\sqrt{1/2m\omega_{1}}$.

In analogy to the EIT system, eq.\eqref{xc2}, we define a reduced mechanical susceptibility $\tilde{\chi}_{M}(\omega_{s})=\rho_{co}(\omega_{s})$. The concept of susceptibility of a mechanical oscillator is widely used in optomechanics \cite{Vahala2007, Kippenberg2008, Kippenberg2010, Painter2011}. Here we are extending this idea to a set of coupled oscillators. By inspection in eqs.\eqref{e5} and \eqref{e9} we see that $\rho_{31}$ and $\rho_{co}$ are perfectly equivalent. Thus, the classical analog of each parameter of EIT in atomic physics can be identified formally in the mechanical system, as summarized in table \ref{TableEIT} and illustrated in Fig.\ref{EsquemaEIT}(c). Each harmonic oscillator is identified as a dipole-allowed transition with electronic dipole moment $\mu_{i3}$ ($i=1,2$).

The classical analog for the two-photon detuning $\delta=\Delta_{p}-\Delta_{c}$ is $\Delta_{2}=\Delta_{1}-\Delta_{21}$, where $\Delta_{21}$ accounts for the detuning of the resonant frequencies between oscillator 2 and oscillator 1. It can be obtained readily by setting $k_{2}=k_{1}\pm\Delta k$. The detuning $\Delta_{21}$ is responsible for reproducing the shift observed in the dark state when $\Delta_{c}\neq0$. The atomic transitions of the EIT system are considered to have fixed resonant frequencies $\omega_{31}$ and $\omega_{32}$, meaning that the detuning $\Delta_{c}$ is performed by changing the frequency of the control field $\omega_{c}$. In the mechanical system the equivalent of $\omega_{c}$ is $\omega_{12}$ but the classical detuning $\Delta_{21}$ is performed by changing the spring constants $k_{1}$ or $k_{2}$ and not $k_{12}$. This is because $\omega_{1}$ and $\omega_{2}$ depends on
$k_{12}$ in the same way. Then we have to keep $\omega_{12}$ constant by fixing $k_{12}$ and change the resonant frequencies $\omega_{1}$ and $\omega_{2}$ through $k_{1}$ and $k_{2}$ to produce the detuning $\Delta_{21}$. For perfect control field resonance $\Delta_{c}=0$, we have $\delta =\Delta_{p}$, which corresponds to $\Delta_{1}=\Delta_{2}$ in eq.\eqref{e9},
implying that $\omega_{1}=\omega_{2}$ and consequently $k_{1}=k_{2}$ for the coupled oscillators.

%\begin{center}
\begin{table}[th]
\caption{Classical analog of EIT using two mechanical coupled harmonic oscillators (2-MCHO).}%
\label{TableEIT}
\centering
\begin{tabular} {c c}
\hline\hline
EIT $\left(\rho_{31}\right)$ & 2-MCHO $\left(\rho_{co}\right)$\\[1ex]
\hline
$\Delta_{p}$ & $\Delta_{1}$\\
$\delta$ & $\Delta_{2}$\\
$\Omega_{p}$ & $\Omega_{s}$\\
$\Omega_{c}$ & $\Omega_{12}$\\
$\gamma_{31}$ & $\gamma_{1}$\\
$\gamma_{2}$ & $\gamma_{2}$\\ [1ex]
\hline
\end{tabular}
\end{table}
%\end{center}

\begin{figure}[!ht]
\includegraphics[width=8.5cm]{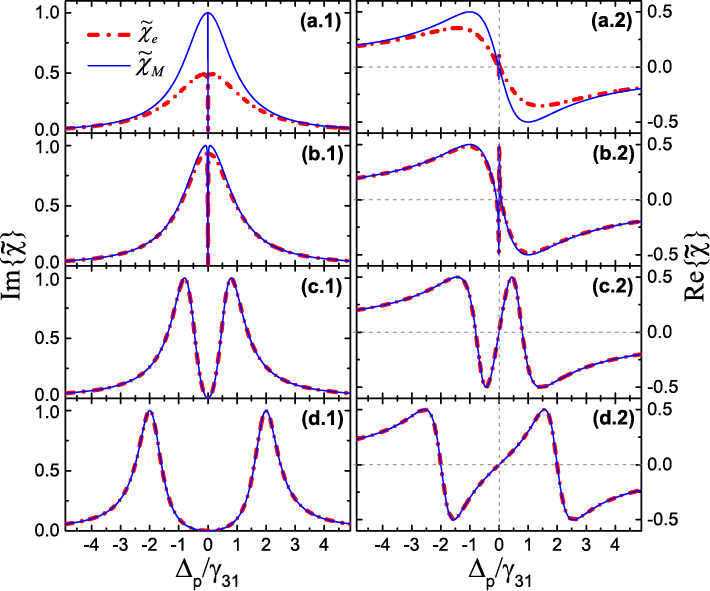}
\caption{(Color online). Imaginary and real parts of the reduced electric susceptibility $\tilde{\chi}_{e}$ vs the normalized probe-atom detuning $\Delta_{p}/\gamma_{31}$ for the EIT system in comparison with its classical counterpart $\tilde{\chi}_{M}$ for $\Omega_{p}=0.02\gamma_{31}$, $\gamma_{2}=0$, $\Delta_{c}=0$ and different values of the Rabi frequency of the
control field (a) $\Omega_{c}=0.02\gamma_{31}$, (b) $0.08\gamma_{31}$, (c) $0.8\gamma_{31}$ and (d) $2.0\gamma_{31}$. For the mechanical system we use the same set of parameters following the analog presented in Table \ref{TableEIT}.} \label{FigCOEIT}
\end{figure}

In Fig.\ref{FigCOEIT} we show the imaginary and real parts of the reduced electric susceptibility $\tilde{\chi}_{e}$ vs the normalized probe-atom detuning $\Delta_{p}/\gamma_{31}$ for the EIT system in comparison with its mechanical counterpart $\tilde{\chi}_{M}$, obtained using two coupled oscillators. The parameters in the classical system are set to be the same as
in the EIT following the analog presented in Table \ref{TableEIT}, for $\Omega_{p} = 0.02\gamma_{31}$, $\gamma_{2} = 0$, $\Delta_{c} = 0$ and different values of the Rabi frequency of the control field $\Omega_{c}$.

For the set of parameters used in Figs.\ref{FigCOEIT}(a) and \ref{FigCOEIT}(b) the EIT condition $\Omega_{p}<<\Omega_{c}$ is not deeply satisfied. Once $\rho_{11}\neq 1$ in these cases, the classical model do not reproduce the atomic result satisfactorily. When the condition is fulfilled, $\rho_{11} \approx 1$, we have perfect equivalence between the classical and semiclassical results, as depicted in Figs.\ref{FigCOEIT}(c) and \ref{FigCOEIT}(d).

If the EIT condition $\Omega_{p}<<\Omega_{c}$ is deeply satisfied the absorption profile of EIT presented in Fig.\ref{FigCOEIT}, for $\gamma_{2} = 0$, remains observable even for nonvanishing $\gamma_{2}$, since the condition $\gamma_{2}<<\gamma_{31}$ is prescribed \cite{Fleischhauer2005}. In this way, the classical model reproduces the atomic system for any set of parameters.

If we recall the dressed states analysis for EIT, the dark state, given by the transparency window observed between the two peaks of absorption, is writen as the superposition between the bare ground states $\left|1\right\rangle$ and $\left|2\right\rangle$ and not the excited state $\left|3\right\rangle$. This means that an atom in this state has no probability of absorbing or emitting a photon. The idea of quantum interference process behind the cancelation of absorption in EIT systems is widely described in the literature \cite{Harris1997, Marangos1998, Fleischhauer2005}. When classical analogies for such systems are presented, like the one we are discussing here, many questions arises as to: What physical property is transparent for the coupled oscillators? What is interfering in this system? And most importantly, what is the classical dark state in this case?

The first question was already responded by Alzar et. al. \cite{Nussenzveig2002}. They showed that the classical observable related to the EIT absorption profile is given by the real part of the average power absorbed by oscillator 1, owing the application of the harmonic force $F_{s}(t)$, while the dispersive behavior is contained in the real part of the frequency dependence of the amplitude of $x_{1}$. Note that these observables are in fully agreement with our definition of the reduced mechanical susceptibility $\tilde{\chi}_{M}(\omega_{s})=\rho_{co}(\omega_{s})$. The power absorbed by oscillator 1 is given by $P_{s}(t) = F_{s}(t)\dot{x}_{1}(t) = -i\omega_{s}F_{s}(t)x_{1}(t)$. The relation between $P_{s}$ and $\rho_{co}$ is drawn from eq.\eqref{e8} through $x_{1}(t)$. Once $P_{s}$ is multiplied by the imaginary number $i$, the imaginary part of $\rho_{co}$ depicted in Fig.\ref{FigCOEIT} is related to the real part of $P_{s}$. Equation \eqref{e8} also provides a straightforward relation between the dispersive behavior, defined by Alzar, and the real part of $\rho_{co}$ in Fig.\ref{FigCOEIT}, once $\rho_{co}$ is contained in the amplitude of $x_{1}$.

In analogy to the dressed states analysis, if we recall the normal modes description for the coupled oscillators system we can answer to the remaining questions readily. All calculations are described in detail in Appendix B. Considering the simplified case where $m_{1,2} = m$, $k_{1,2} = k$ and the definition of the normal coordinates $X_{+}$ and $X_{-}$, which are linear combinations of $x_{1}(t)$ and $x_{2}(t)$, the coupled Hamiltonian \eqref{HamiltA}, described in appendix A, can be written as a combination of two uncoupled forced harmonic oscillators with normal resonance frequencies $\omega_{+} = \sqrt{k/m}$ and $\omega_{-} = \sqrt{\omega^{2}_{+} + 2\omega^{2}_{12}}$. These are the resonance frequencies of the two normal modes of the system, usually named as the symmetric $NM_{(+)}$ and asymmetric $NM_{(-)}$ modes. In $NM_{(+)}$ both masses move in exactly the same way, meaning that the middle spring is never stretched, while in $NM_{(-)}$ the masses move oppositely. This means that any arbitrary motion of the system, like the displacement of oscillator 1 or 2, is a linear combination of those two normal modes. In other words, $x_{1,2}(t)$ can be seen as a superposition of two harmonic motions.

The EIT-like profile is observed when the damping forces are considered. In this case the eqs.\eqref{NCX} in Appendix B for the normal modes, become coupled through the damping constants $\gamma_{1}$ and $\gamma_{2}$. Solving for the steady-state solution we find a relationship between the normal coordinates $X_{+}$ and $X_{-}$ which depends on the frequencies of the normal modes $\omega_{\pm}$, the frequency of the force $\omega_{s}$ and the damping constant of oscillator 2, $\gamma_{2}$, see eq.\eqref{relNM}. As we are probing the response of oscillator 1 due to the harmonic force $F_{s}(t)$, the classical dark state is observed when $\omega_{s} = \omega_{1}$, with $\omega_{1}^{2} = \omega_{+}^{2} + \omega_{12}^{2}$. Note that the frequency $\omega_{1}$ sits in the range between $\omega_{+}$ and $\omega_{-}$, which is a region of high probability to occurs interference between the normal modes.

As we have discussed the EIT transparency window, which characterizes the dark state, is observed when the conditions $\Omega_{p}<<\Omega_{c}$ and $\gamma_{2}<<\gamma_{31}$ are prescribed. According to Table \ref{TableEIT} the classical analog for these conditions are $\Omega_{s}<<\Omega_{12}$ and $\gamma_{2}<<\gamma_{1}$. Considering $\gamma_{2} \rightarrow 0$, as in Fig.\ref{FigCOEIT}, and $\omega_{s} = \omega_{1}$ we find that $X_{+} = - X_{-}$. Once the displacement of oscillators 1 and 2 are given by $x_{1,2} = \sqrt{2}/2\left(X_{+} \pm X_{-}\right)$, we have $x_{1} = 0$ and $x_{2} \neq 0$ in this case. From eq.\eqref{e8} $x_{1} = 0$ is fulfilled for $\rho_{co} = 0$, as observed in Fig.\ref{FigCOEIT} for zero detuning. Thus, the classical dark state is obtained when oscillator 1 stays stationary while oscillator 2 oscillates harmonically. In other words, oscillator 1 becomes transparent to the effect of the driving force for $\omega_{s} = \omega_{1}$ conducting to zero power absorption, which is a consequence of a destructive interference between the two normal modes $NM_{(\pm)}$ in the displacement of oscillator 1.

The first classical condition $\Omega_{s}<<\Omega_{12}$ becomes necessary for small but non-zero $\gamma_{2}$, i.e., $\gamma_{2}<<\gamma_{1}$. In this case the classical dark state remains observable for $k_{12}>>k_{1}$, meaning that $X_{+} \approx - X_{-}$, see eq.\eqref{Nmodes} in Appendix B. From the definitions of $\Omega_{s}$ and $\Omega_{12}$ one can find readily that $\Omega_{s} = F\sqrt{\Omega_{12}/k_{12}}$, showing that the relation between $\Omega_{s}$ and $\Omega_{12}$ also depends on $F$ as expected. Similarly to the atomic system, where the probe field is turned on slowly for the state $\left|1\right\rangle$ evolving into the dark state and decouple from the other states, in the classical system the strength of the force, given by the amplitude $F$, is also very small to guarantee the usual approximation of small oscillations. Then, if $F$ is relatively small and $k_{12}>>k_{1}$ we have the condition $\Omega_{s}<<\Omega_{12}$ for nonvanishing $\gamma_{2}$ but $\gamma_{2}<<\gamma_{1}$. Thus, the conditions to observe the phenomenology of EIT can be completely mapped onto the classical system composed by two coupled damped harmonic oscillators, showing that Im$\left\{\tilde{\chi}_{e}(\omega_{p})\right\}\equiv$ Im$\left\{\tilde{\chi}_{M}(\omega_{s})\right\}$ and Re$\left\{\tilde{\chi}_{e}(\omega_{p})\right\}\equiv$ Re$\left\{\tilde{\chi}_{M}(\omega_{s})\right\}$ since $\Omega_{p}<<\Omega_{c}$ and $\gamma_{2}<<\gamma_{31}$.

The similarities obtained between the EIT atomic system and the mechanical coupled oscillators are not surprising. Many aspects of the atom-field interaction can be described by the classical theory of optical dispersion \cite{Christy1972, Lipson2011}. According to this theory systems which can be approximated by two discrete levels are represented as classical harmonic oscillators. Then, the classical picture of a two-level atomic system consists of a massive positive nucleus surrounded by an electron cloud with an equal negative charge. The electron of charge $q$ and mass $m$ is supposed to be bound to the immovable nucleus by a linear restoring force $-kx$, where $x$ is the distance between their centres of mass and charge. For the static case these centres are coincident and the atom has zero dipole moment. The energy loss is introduced phenomenologically as a damping force proportional to
velocity $-\eta\dot{x}$. If the atom is disturbed by an electromagnetic field $E$, there is also an applied force on the electron $F_{q}=qE$, and then, the electron cloud oscillates along the centre of mass. Thus, we have an oscillating dipole with dynamics described by the same equation of motion of a forced, damped harmonic oscillator, $m\ddot{x}+\eta\dot{x}+kx=F_{q}$, which is the same obtained previously for the first oscillator if $k_{12}=0$. Once the EIT phenomenon is observed in an ensemble of noninteracting three-level atoms in their ground states, it provides an instructive example of the extension of the classical theory of optical dispersion for multi-level systems. Each atomic transition behaves as a harmonic oscillator which loses energy by some mechanical friction mechanism.

If we turn back to the physical analogy between EIT and the classical model reported by Alzar et al. \cite{Nussenzveig2002}, the atom is represented by oscillator 1. According to the classical theory presented previously, this would be correct if the atom has two discrete levels of energy, i.e., only one dipole-allowed transition, which is not the case. As we are dealing with three-level atoms, the correct is to represent each dipole transition as a harmonic oscillator. According to the classical picture for the atom, displayed in Fig.\ref{EsquemaEIT}(c), the dipole transition frequencies $\omega_{31}$ and $\omega_{32}$ correspond to the natural frequencies of particles 1 and 2, respectively. The analog for the control and probe fields are equivalent to those presented in \cite{Nussenzveig2002}, where they are identified by the coupling spring and by the harmonic force acting on particle 1, respectively.

For other classical systems, like RLC coupled circuits and acoustic structures, the analog of the EIT absorption is also obtained from the real part of the power absorbed by the pumped oscillator \cite{Nussenzveig2002, Tokman2002, Huang2010, Serna2011, Huang2013}.

In what follows the classical analog for different quantum systems are presented using the same configuration for the two mechanical coupled harmonic oscillators model discussed here.

\subsection{EIT-like in two coupled optical cavities}

Once we can reproduce the phenomenology of EIT with two classical coupled oscillators it is natural to consider the oscillators quantum mechanically and see the consequences of it in the EIT-like phenomenon and its conditions \cite{Ponte2005}. For this end, we used a model consisting of two coupled optical cavities with one of them pumped by a coherent field. The use of optical cavities is convenient because we will show the classical analog for EIT-related phenomena in systems comprised by a single two- or three-level atom coupling a single mode of an optical resonator.

\begin{figure}[!ht]
\includegraphics[width=6cm]{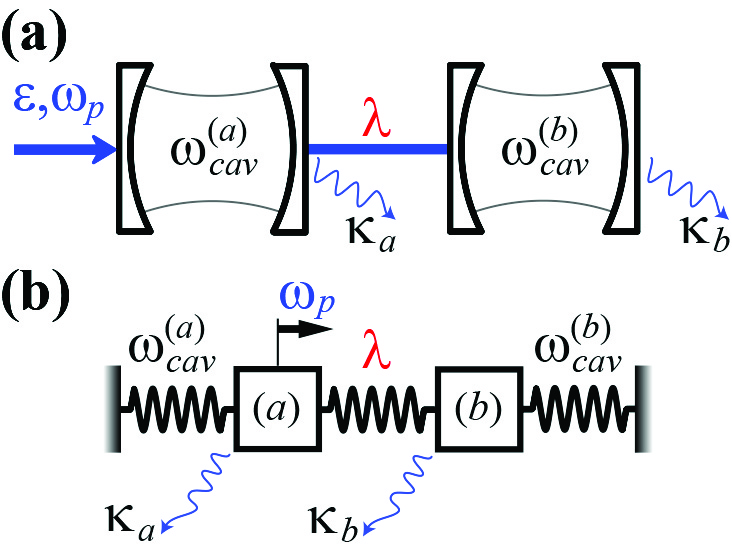}
\caption{(Color online). (a) Two coupled cavities showing their respective single cavity modes with frequencies $\omega_{cav}^{(a)}$, $\omega_{cav}^{(b)}$ and cavity decay rates $\kappa_{a}$, $\kappa_{b}$. Cavity $A$ is driven by a classical probe field with frequency $\omega_{p}$ and strength $\varepsilon$. The electromagnetic modes exchange energy with Rabi frequency $2\lambda$. (b)
Classical analog showing the equivalence for each parameter of the coupled cavity modes in the mechanical system.} \label{CoupledCavity}
\end{figure}

The two single electromagnetic modes of frequencies $\omega_{cav}^{(a)}$ and $\omega_{cav}^{(b)}$ of optical resonators $a$ and $b$, respectively, exchange energy with Rabi frequency $2\lambda$. Cavity $a$ is driven by a coherent field (probe) of frequency $\omega_{p}$ and strength $\varepsilon$, as illustrated in Fig.\ref{CoupledCavity}(a).

Introducing the rotating-wave approximation (RWA) and considering identical frequencies $\omega_{cav}^{(a)}=\omega_{cav}^{(b)}=\omega_{cav}$ for simplicity, the time-independent Hamiltonian which describes the cavity-cavity coupling in the probe laser rotating frame is given by
\begin{equation}
H=\Delta_{cav}\left(a^{\dagger}a+b^{\dagger}b\right) + \lambda\left(ab^{\dagger}+a^{\dagger}b\right)  +\varepsilon\left(  a+a^{\dagger}\right).\label{e10}
\end{equation}
Since the cavity modes are quantized, they are expressed in terms of creation ($a^{\dagger},b^{\dagger}$) and annihilation ($a,b$) operators. $\Delta_{cav}=\omega_{cav}-\omega_{p}$ is the probe-cavity detuning. The master equation for the cavity-cavity density operator is
\begin{equation}
\dot{\rho}=-i[H,\rho]+\sum\limits_{\alpha=a,b}\kappa_{\alpha}(2\alpha \rho\alpha^{\dagger}-\alpha^{\dagger}\alpha\rho-\rho\alpha^{\dagger}\alpha) \label{e11}
\end{equation}
where $\kappa_{\alpha}$ is the cavity mode decay rate of cavity $\alpha$. The time evolution for the expectation value of the field operators are
\begin{subequations}\label{EvoCav}
\begin{align}
\left\langle \dot{a}\right\rangle &= -i\left\{ \left(\Delta_{cav} - i\kappa_{a}\right)\left\langle a\right\rangle + \lambda\left\langle b\right\rangle + \varepsilon \right\},\\
\nonumber\\
\left\langle \dot{b}\right\rangle &= -i\left\{ \left(\Delta_{cav} - i\kappa_{b}\right)\left\langle b\right\rangle + \lambda\left\langle a\right\rangle \right\},
\end{align}
\end{subequations}\label{}
which exhibits essentially the same structure as eqs.\eqref{HRhosB}, Appendix A, for $\dot{\rho}_{\alpha}$ and $\dot{\rho}_{\beta}$, respectively, in the description of the dynamics of the coupled oscillators system.

Once the cavity mode $a$ absorbs photons from the pumping field and communicates them to cavity $b$, through the coupling $\lambda$, we represent the probe response of the cavity-cavity system as a reduced electric susceptibility given by the expectation value of the driven cavity field, i.e., $\tilde{\chi}_{CC}(\omega_{p})=\left\langle a\right\rangle $. Note that it is precisely what was done for the EIT medium, where $\tilde{\chi}_{e}(\omega_{p})=\rho_{31}$. A formal correspondence between $\rho_{31}$ in atomic physics and the intracavity field $\left\langle a\right\rangle$ was already pointed out by Stefan Weiss et al. in their work about optomechanically induced transparency \cite{Kippenberg2010}.

The steady state solutions of the expectation value of field operators in eqs.\eqref{EvoCav} provide the solution for the intracavity field of cavity $a$:
\begin{equation}
\left\langle a\right\rangle =\frac{-\varepsilon\left(  \Delta_{cav}-i\kappa_{b}\right)}{\left(\Delta_{cav}-i\kappa_{a}\right)  \left(\Delta_{cav}-i\kappa_{b}\right)  -\lambda^{2}}, \label{e12}
\end{equation}
which is identical to the reduced mechanical susceptibility $\tilde{\chi}_{M}(\omega_{s})=\rho_{co}$ obtained for the two coupled harmonic oscillators in eq.\eqref{e9} for $\Delta_{1}=\Delta_{2}=\Delta_{s}$. The negative signal observed in eq.\eqref{e12} can be reproduced from the classical equations by considering the phase $\phi_{s}=\pi$ in the applied force on oscillator 1, which is equivalent to make $-F$ in eq.\eqref{e9}. Once it is considered only one force in the classical analog the phase is not relevant. Nonetheless it becomes important for atomic systems with more than three-levels, like in the four-level tripod configuration we show afterwards, in which the classical analog is obtained by considering two oscillating forces out of phase by $\pi$.

The classical analog of each parameter of the coupled cavity modes is summarized in table \ref{TableEITCC}. The cavity EIT-like condition is given by $\varepsilon<< \lambda$ and $\kappa_{b} << \kappa_{a}$ and the classical analog is obtained for any set of parameters.

\begin{table}[th]
\caption{Classical analog of EIT-like in two coupled cavity modes (EIT-CCM) using two mechanical coupled harmonic oscillators (2-MCHO).}
\label{TableEITCC}
\centering
\begin{tabular} {c c}
\hline\hline
EIT-CCM $\left(\left\langle a\right\rangle \right)  $ & 2-MCHO $\left(\rho_{co}\right)$ \\[1ex]
\hline
$\Delta_{cav}$ & $\Delta_{s}$\\
$\varepsilon$ & $\Omega_{s}$\\
$\lambda$ & $\Omega_{12}$\\
$\kappa_{a}$ & $\gamma_{1}$\\
$\kappa_{b}$ & $\gamma_{2}$ \\[1ex]
\hline
\end{tabular}
\end{table}

The agreement between the cavity-field and oscillator-force responses is somehow expected. In the quantum theory of radiation \cite{Scully1997} a general multimode field is represented by a collection of harmonic oscillators, one for each mode. Then, the single mode of the electromagnetic field of cavity $a$ or $b$ is dynamically equivalent to a simple harmonic
oscillator. Once we have two coupled cavity modes, naturally it will be equivalent to two coupled oscillators.

Narducci \textit{et al.} \cite{Narducci1968} showed that differences in the dynamics of two coupled quantum oscillators may arise between the approximated Hamiltonian given by eq.\eqref{e10} and its exact solution, where the counter rotating-wave terms $a^{\dagger}b^{\dagger}$ and $ab$ are considered. They established the limits of validity of the RWA in terms of the strength of coupling $\lambda$. Our results show that, if the RWA is assumed to be valid, the quantum dynamics of two coupled cavity modes can be reproduced by the classical dynamics of two coupled harmonic oscillators. Thus, to obtain the classical analog for systems which involve a cavity mode, we can represent it as a harmonic oscillator with natural frequency $\omega_{cav}$, similarly to an atomic dipole-allowed transition in the low atomic excitation condition.

The result obtained in eq.\eqref{e12} goes beyond than the perfect agreement between quantum and classical models. It opens the possibility of a physical interpretation for the expectation value of the photon annihilation operator $\left\langle a\right\rangle$, showing that it is directly related to the electric susceptibility of a cavity mode. In what follows we show that this interpretation can also be used for systems comprised by two- and three-level atoms interacting with a single cavity mode driven by a coherent field.

\subsection{EIT-like in two-level atom coupled to an optical cavity mode}

The absorption spectrum of EIT is also observed when a single two-level atom is coupled to a single cavity mode. This effect was predicted by Rice and Brecha \cite{Brecha1996} and termed as cavity induced transparency (CIT). They found that under specific conditions an atom-cavity transmission window, usually referred to as intracavity dark state, arises as a consequence of quantum interference between two absorption paths and not as a result of vacuum-Rabi splitting. They showed the analogous in the weak-probe limit considering the driven cavity and the driven atom cases. We will examine both configurations and show their classical equivalent using two coupled oscillators.

First we consider the driven cavity case. The system is comprised of a single atom with two energy levels, $\left\vert g\right\rangle$ and $\left\vert e\right\rangle$, coupled to a single electromagnetic mode of frequency $\omega_{cav}$ of an optical resonator. The cavity is driven by a coherent field (probe) with frequency $\omega_{p}$ and strength $\varepsilon_{c}$. The atomic transition $|g\rangle\leftrightarrow|e\rangle$ (frequency $\omega_{0}$) is coupled by the cavity mode with vacuum Rabi frequency $2g$. The time-independent Hamiltonian which describes the atom-field coupling in a rotating frame is obtained using the driven Jaynes-Cummings model
\begin{equation}
H = \Delta_{0}\sigma_{ee}+\Delta_{c}a^{\dagger}a+g\left(a\sigma_{eg}+a^{\dagger}\sigma_{ge}\right) + \varepsilon_{c}\left(a+a^{\dagger}\right), \label{e13}
\end{equation}
with detunings given by $\Delta_{0}=\omega_{0}-\omega_{p}$, $\Delta_{c}=\omega_{cav}-\omega_{p}$.

The master equation for the atom-cavity density operator is
\begin{align} \label{e14}
\dot{\rho} & = -i[H,\rho] + \kappa(2a\rho a^{\dagger} - a^{\dagger}a\rho - \rho a^{\dagger}a) \nonumber\\
& + \Gamma_{eg}(2\sigma_{ge}\rho\sigma_{eg} - \sigma_{ee}\rho - \rho\sigma_{ee}) \nonumber\\
& + \gamma_{e}(2\sigma_{ee}\rho\sigma_{ee} - \sigma_{ee}\rho - \rho\sigma_{ee}),
\end{align}
where $\kappa$ is the cavity-field decay rate, $\Gamma_{eg}$ the polarization decay rate of the excited level $|e\rangle$ to the level $|g\rangle$, and $\gamma_{e}$ the non-radiative atomic dephasing rate of state $|e\rangle$. By using the commutation relation $\left[  a,a^{\dagger}\right]  =1$ and considering perfect atom-cavity resonance $\omega_{0}=\omega_{cav}$, implying that $\Delta_{0}=\Delta_{c}$, the time evolution of the expected values of the atomic and field operators are given by
\begin{subequations}\label{e15}
\begin{align}
\left\langle \dot{a}\right\rangle &= -i\left\{ \left(\Delta_{c} - i\kappa\right)\left\langle a\right\rangle + g\left\langle \sigma_{ge}\right\rangle + \varepsilon_{c} \right\},\label{e15a}\\
\left\langle \dot{\sigma}_{ge}\right\rangle &= -i\left\{ \left(\Delta_{c} - i\gamma_{eg}\right) \left\langle \sigma_{ge}\right\rangle - g\left\langle a\right\rangle \left\langle \sigma_{z} \right\rangle \right\}, \label{e15b}
\end{align}
\end{subequations}
where $\gamma_{eg}=\Gamma_{eg}+\gamma_{e}$ and $\left\langle \sigma_{z}\right\rangle =\left\langle \sigma_{ee}\right\rangle -\left\langle\sigma_{gg}\right\rangle $.

The closed set of coupled equations above are obtained by using a semiclassical approximation \cite{Ren1994}, which consists of factoring joint operator moments $\left\langle a\sigma\right\rangle \rightarrow\left\langle a\right\rangle \left\langle \sigma\right\rangle $. Thereby, the cavity field is described by a complex amplitude $\left\langle a\right\rangle = \alpha$ rather than a quantum mechanical operator.

The EIT-like phenomenon in this system is observed when the Rabi frequency of the cavity field $g\left\langle a\right\rangle _{max}$ is large compared to the Rabi frequency of the probe field, $\varepsilon_{c}<<g\left\langle a\right\rangle _{max}$, and also when $\gamma_{eg}<<\kappa$. The average $\left\langle a\right\rangle_{max}=\varepsilon_{c}/\left(  \Delta_{c}-i\kappa\right)$ is the maximum value of $\left\langle a\right\rangle$ in the absence of atoms $(g=0)$. As we have seen in the previous section, the optical response of the atom-cavity medium is proportional to the expectation value of the cavity field $\left\langle a\right\rangle $, once the cavity mode is pumped weakly by the probe field. Then, we will represent the probe response as an atom-cavity reduced susceptibility $\tilde{\chi}_{AC}(\omega_{p})=\left\langle a\right\rangle $. The real part of $\tilde{\chi}_{AC}$ is related to the absorption spectrum of the system and its imaginary part to the phase of the outgoing light field of the cavity. In the steady state, $\dot{\rho}=0$, the equations above give for the expectation value of the photon annihilation operator,
\begin{equation}
\left\langle a\right\rangle =\frac{-\varepsilon_{c}\left(\Delta_{c}-i\gamma_{eg}\right)}{\left(\Delta_{c}-i\kappa\right)  \left(\Delta_{c}-i\gamma_{eg}\right) + g^{2}\left\langle \sigma_{z}\right\rangle}.\label{e16}
\end{equation}

\begin{figure}[!ht]
\includegraphics[width=8cm]{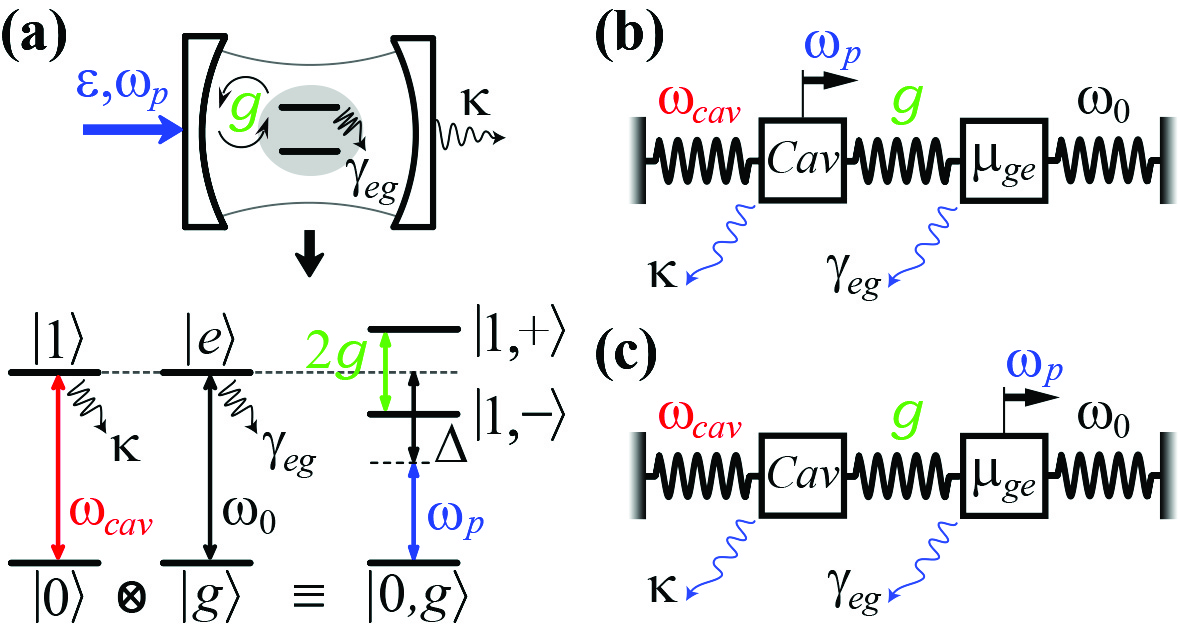}
\caption{(Color online). (a) \textit{Top:} Single two-level atom with resonance frequency $\omega_{0}$ and atomic polarization decay rate $\gamma_{eg}$, interacting with a single mode of an optical resonator with frequency $\omega_{cav}$ and cavity decay rate $\kappa$. The atom-field dipole coupling is described by the vacuum Rabi frequency $2g$. A classical probe field with frequency $\omega_{p}$ and strength $\varepsilon$ pumps either the cavity or the atom. \textit{Bottom:} First doublet of dressed-states of the Jaynes-Cummings ladder as a result of the coupling between the bare cavity ($\left\vert 0\right\rangle ,\left\vert 1\right\rangle $) and the bare atom ($\left\vert g\right\rangle ,\left\vert e\right\rangle $). (b) and (c) show the atom-field classical analogs for the driven cavity and driven atom cases, respectively.} \label{AtomCavity}
\end{figure}

If $\left\langle \sigma_{z}\right\rangle =-1$, $\left\langle a\right\rangle$ becomes identical to the reduced mechanical susceptibility $\tilde{\chi}_{M}(\omega_{s})=\rho_{co}$, see eq.\eqref{e9}. Mathematically, $\left\langle\sigma_{z}\right\rangle =-1$ is the limit to reach low atomic excitation, meaning that the probe field is so weak that we can consider only the zero- and one-photon states ($\left\vert 0\right\rangle ,\left\vert 1\right\rangle$) of the cavity mode. As illustrated in Fig.\ref{AtomCavity}(a), the atom-field system will be limited to the first splitting of the dressed states which forms the anharmonic Jaynes-Cummings ladder.

The atom-field classical analog for the driven cavity case is shown in Fig.\ref{AtomCavity}(b) and each parameter is identified as in table \ref{ACsystems}. It is also interesting to make comparisons between the original EIT-$\Lambda$ scheme and other quantum systems. In this case, the cavity makes the role of the atomic transition $|1\rangle\leftrightarrow
|3\rangle$ and the atom represents the transition $|2\rangle\leftrightarrow |3\rangle$, see Figs.\ref{EsquemaEIT}(a) and \ref{EsquemaEIT}(c).

\begin{figure}[!ht]
\includegraphics[width=8.5cm]{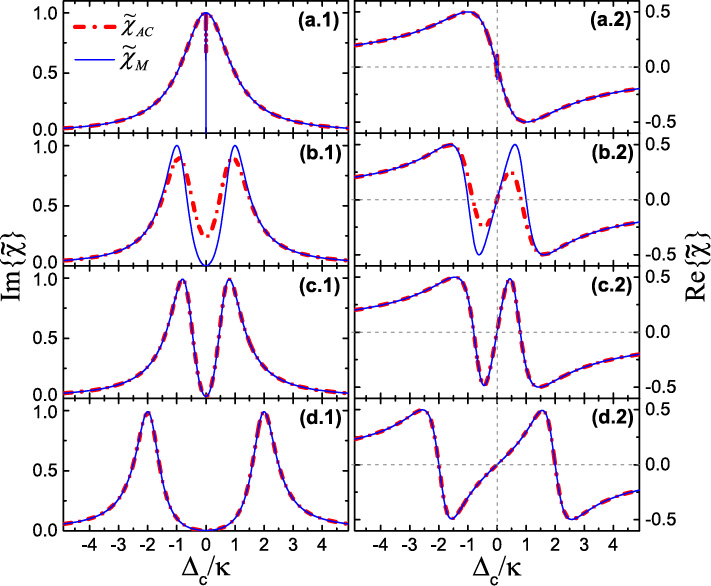}
\caption{(Color online). Imaginary and real parts of the reduced atom-cavity susceptibility $\tilde{\chi}_{AC}$ vs the normalized probe-cavity detuning $\Delta_{c}/\kappa$ for the two-level atom interacting with a single mode of a driven optical cavity in comparison with its mechanical analog $\tilde{\chi}_{M}$. The parameters are (a) $\varepsilon_{c}=0.02\kappa$, $g=0.02\kappa$, $\gamma_{eg}=0.0$, (b) $0.5\kappa$, $1.0\kappa$, $0.0$, (c) $0.02\kappa$, $0.8\kappa$, $0.01\kappa$ and (d) $0.02\kappa$, $2.0\kappa$, $0.01\kappa$. The classical results were obtained using the same set of parameters following the analog depicted in table \ref{ACsystems}.} \label{TLPC}
\end{figure}

Figure \ref{TLPC} shows the imaginary and real parts of the reduced susceptibility $\tilde{\chi}_{AC}(\omega_{p})$ vs the normalized probe-cavity detuning $\Delta_{c}/\kappa$ for different set of parameters in comparison with its classical analog $\tilde{\chi}_{M}(\omega_{s})$. The full quantum atom-cavity description is solved for the steady state of $\rho$ following the method presented in \cite{Tan1999}, where the cavity field Fock basis is truncated according to the probe strength.

In Figs.\ref{TLPC}(a) and \ref{TLPC}(b) the EIT-like condition $\varepsilon_{c}<<g\left\langle a\right\rangle_{max}$ is not deeply satisfied, showing that the intracavity dark-state $\left\langle a\right\rangle =0$ for $\Delta_{c}=0$ is not observed, differently for its classical counterpart once $\gamma_{eg} \equiv \gamma_{2} = 0$, see Appendix B. When such condition is fulfilled the results show perfect agreement even for nonvanishing $\gamma_{eg}$, like in Figs.\ref{TLPC}(c) and \ref{TLPC}(d), since $\gamma_{eg}<<\kappa$.

As we have mentioned the condition $\left\langle \sigma_{z}\right\rangle =-1$ in eq.\eqref{e16} means the atom-cavity field can be described by the first doublets of dressed-states of the Jaynes-Cummings ladder, see Fig.\ref{AtomCavity}(a), regardless the atom-cavity system being considered in the strong coupling regime $g>>(\gamma_{eg},\kappa)$, like in Fig.\ref{TLPC}(d). Thus, the quantum atom-field correlations can be completely neglected and then, atom and cavity field can be treated in the same footing as harmonic oscillators. In ref.\cite{Rempe2014} the authors used the full classical result, given by eq.\eqref{e16}, to analyze experimentally the measurement of antiresonances in a strongly-coupled atom-cavity system by using heterodyne detection.

The aspects of EIT-like phenomenon regarding the spectrum of absorption obtained from the imaginary part of $\left\langle a\right\rangle $, can also be observed through the calculation of cavity transmission. It is provided by the average photon number $\left\langle a^{\dagger}a\right\rangle $. Once we have the classical analog for $\left\langle a\right\rangle \equiv\rho_{co}$, one can see readily that $\left\langle a^{\dagger}a\right\rangle \equiv \rho^{\ast}_{co}\rho_{co}$.

For the driven atom case, the probe field with strength $\varepsilon_{0}$ pumps the atom instead of the cavity mode. For this system, the time-independent Hamiltonian in a rotating frame reads
\begin{equation}
H = \Delta_{0}\sigma_{ee}+\Delta_{c}a^{\dagger}a+g\left(  a\sigma_{eg}+a^{\dagger}\sigma_{ge}\right)  +\varepsilon_{0}\left(\sigma_{eg}+\sigma_{ge}\right). \label{e17}
\end{equation}

As before we consider atom and cavity on resonance $\omega_{0}=\omega_{cav}$, then $\Delta_{c}=\Delta_{0}$, where $\Delta_{0}=\omega_{0}-\omega_{p}$ is the probe-atom detuning. Once the probe field couples directly to the atom, the probe absorption is related to the density matrix element $\rho_{eg}=\left\langle \sigma_{ge}\right\rangle$, in analogy with $\rho_{31}$ in eq.\eqref{SFull}. Then, the atom-cavity reduced susceptibility is represented by $\tilde{\chi}_{AC}(\omega_{p})=\left\langle \sigma_{ge}\right\rangle $. Using the master equation (\ref{e14}) to obtain the time evolution for the atomic and field operators, we solve for the expectation value of the lowering atomic operator in the steady state,
\begin{equation}
\left\langle \sigma_{ge}\right\rangle =\frac{\varepsilon_{0}\left\langle\sigma_{z}\right\rangle \left(  \Delta_{0}-i\kappa\right)}{\left(\Delta_{0}-i\gamma_{eg}\right) \left(\Delta_{0}-i\kappa\right)+g^{2}\left\langle \sigma_{z}\right\rangle }, \label{ACPA}
\end{equation}
which is also identical to the mechanical reduced susceptibility $\tilde{\chi}_{M}=\rho_{co}$ for $\left\langle \sigma_{z}\right\rangle =-1$. Note that eq.\eqref{e16} can be recovered from eq.\eqref{ACPA} by changing $\gamma_{eg}\leftrightarrow\kappa$. Thus, the first EIT-like condition $\varepsilon_{0}<<g\left\langle a\right\rangle_{max}$ remains the same and the second is now switched to $\kappa<<\gamma_{eg}$. The classical analog for this system is illustrated in Fig.\ref{AtomCavity}(d) and each atom-cavity parameter is identified classically in table \ref{ACsystems}.

Differently from Figs.\ref{TLPC}(a) and \ref{TLPC}(b), the dark state is observed in the driven atom for both, classical and quantum responses. Like in the original EIT configuration presented in Fig.\ref{FigCOEIT}, the maximum absorption peaks in the quantum system decreases when the condition $\varepsilon_{0} << g\left\langle a\right\rangle_{max}$ is not deeply
satisfied, meaning that the approximation $\left\langle \sigma_{z}\right\rangle = -1$ is not valid.

The dissipative rates $\gamma_{eg}$ and $\kappa$ for the driven cavity ($\gamma_{eg} << \kappa$) and driven atom ($\kappa<< \gamma_{eg}$) cases, respectively, make the role of the non-radiative atomic dephasing rate of state $|2\rangle$, $\gamma_{2}$, in the EIT system. If those parameters are relatively large the intracavity dark state will be no longer perfect
\cite{Fleischhauer2005}.

Next sections are dedicated to show the classical analog for atomic systems with more than three-levels of energy using three coupled harmonic oscillators.

\begin{table}[ptb]
\caption{Classical analog of EIT for different quantum systems using two mechanical coupled harmonic oscillators (2-MCHO). We present the analogs for the three-level atom in $\Lambda$ configuration (EIT-$\Lambda$), two-coupled cavity modes (EIT-CCM) and two-level atom-cavity systems for the driven cavity (EIT-DC) and driven atom (EIT-DA) cases.}
\label{ACsystems}
\begin{center}
\begin{tabular}{c c c c c}
\hline\hline
EIT-$\Lambda$ & EIT-CCM & EIT-DC & EIT-DA & 2-MCHO\\
$\rho_{31}$ & $\left\langle a\right\rangle $ & $\left\langle a\right\rangle$ & $\left\langle \sigma_{ge}\right\rangle$ & $\rho_{co}$ \\[1ex]
\hline
$\Delta_{p}$ & $\Delta_{p}$ & $\Delta_{c}$ & $\Delta_{0}$ & $\Delta_{s}$\\
$\Omega_{p}$ & $\varepsilon$ & $\varepsilon_{c}$ & $\varepsilon_{0}$ & $\Omega_{s}$\\
$\Omega_{c}$ & $\lambda$ & $g$ & $g$ & $\Omega_{12}$\\
$\gamma_{31}$ & $\kappa_{a}$ & $\kappa$ & $\gamma_{eg}$ & $\gamma_{1}$\\
$\gamma_{2}$ & $\kappa_{b}$ & $\gamma_{eg}$ & $\kappa$ & $\gamma_{2}$ \\[1ex]
\hline
\end{tabular}
\end{center}
\end{table}

\section{Classical analog of EIT in different physical systems using three-coupled harmonic oscillators}

\begin{figure}[!ht]
\includegraphics[width=5cm]{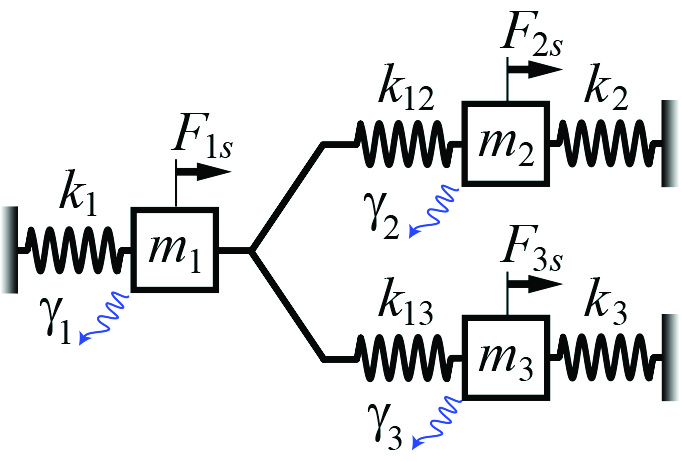}
\caption{(Color online). Mechanical model comprised by three coupled damped harmonic oscillators used to reproduce the EIT-related phenomenology observed in multi-level atomic systems. It consists of three masses $m_{1}$, $m_{2}$ and $m_{3}$ attached to five springs with constant springs $k_{1}$, $k_{2}$, $k_{3}$ for the outside springs and $k_{12}$, $k_{13}$ for the coupling springs. For the general case, a driving force $F_{js}(t)$ of frequency $\omega_{s}$ acts on mass $m_{j}$ and the damping constant of the $j$th harmonic oscillator is represented by $\gamma_{j}$ ($j=1,2,3$).} \label{TMHCO}%
\end{figure}

Now we show how to represent mechanically the EIT-related phenomena observed in four-level atoms in the inverted-Y, tripod and cavity EIT configurations. As we are adding an atomic allowed transition, coupled by a laser field, to the original atomic three-level EIT system, we have to add their classical equivalent in the mechanical system. Then, the mechanical configuration is now composed by three coupled harmonic oscillators as shown in Fig.\ref{TMHCO}.

Hereafter we will follow the same reasoning and notation used for the two coupled oscillators described previously. Considering the general case, where each particle is driven by a coherent force $F_{js}(t)=F_{j}e^{-i(\omega_{s}t+\phi_{s})}+c.c.$ $(j=1,2,3)$ and assuming the solutions $x_{j}=N_{j}e^{-i\omega_{s}t}+c.c.$, the equations of motion on the three masses
give rise to the following equations:
\begin{subequations}
\label{CETL}%
\begin{align}
\left(-\omega_{s}^{2}+\omega_{1}^{2}-2i\gamma_{1}\omega_{s}\right)N_{1}-\omega_{12}^{2}N_{2}-\omega_{13}^{2}N_{3}  &  = \frac{F_{1}}{m}e^{-i\phi_{1}}, \label{CETLA}\\
\left(-\omega_{s}^{2}+\omega_{2}^{2}-2i\gamma_{2}\omega_{s}\right)N_{2}-\omega_{12}^{2}N_{1}  & = \frac{F_{2}}{m}e^{-i\phi_{2}}, \label{CETLB}\\
\left(-\omega_{s}^{2}+\omega_{3}^{2}-2i\gamma_{3}\omega_{s}\right)N_{3}-\omega_{13}^{2}N_{1}  & = \frac{F_{3}}{m}e^{-i\phi_{3}}, \label{CETLC}
\end{align}
\end{subequations}
where $\omega_{1}^{2}=\left(k_{1}+k_{12}+k_{13}\right)/m$, $\omega_{2}^{2}=\left(k_{2}+k_{12}\right)/m$, $\omega_{3}^{2}=\left(k_{3}+k_{13}\right)/m$, $\omega_{12}^{2}=k_{12}/m$, $\omega_{13}^{2}=k_{13}/m$ and $\phi_{j}$ ($j=1,2,3$) the respective phases. As before we consider identical masses $m_{1}=m_{2}=m_{3}=m$ and frequencies $\omega_{j}$ ($j=1,2,3$) near to $\omega_{s}$, implying that the approximations $\omega_{j}^{2}-\omega_{s}^{2}\approx2\omega_{j}(\omega_{j}-\omega_{s})$ and $\gamma_{j}\omega_{s} \approx \gamma_{j}\omega_{j}$ can be used and the corresponding detunings $\Delta_{j}=\omega_{j}-\omega_{s}$ properly defined. As before we have omitted the complex conjugate solution ($c.c.$) for simplicity.

The mechanical representation of the atomic systems we are about to show are more complicated owing the amount of dipole transitions and coupling fields. Depending on the atomic configuration, we will choose which particle or particles in the classical system are driven by the corresponding forces $F_{js}(t)$.

The collective motion of the system for the configuration presented in Fig.\ref{TMHCO} is described by three normal modes, owing the addition of the third mass. Considering the simple case, where $k_{i} = k$ ($i=1,2,3$) and $k_{1j} = k_{\alpha}$ with $\omega_{1j}^{2} = \omega^{2} = k_{\alpha}/m$ ($j=2,3$), the resonance frequencies are $\omega_{0} = \sqrt{k/m}$, $\omega_{+} = \sqrt{\omega_{0}^{2} + \omega^{2}}$ and $\omega_{-} = \sqrt{\omega_{0}^{2} + 3\omega^{2}}$, which are the frequencies of the normal modes $NM_{(0)}$, $NM_{(+)}$ and $NM_{(-)}$, respectively. The modes $NM_{(0)}$ and $NM_{(-)}$ are similar to the two normal modes described in Sec.\ref{SecEIT}. In $NM_{(0)}$ the three masses move in phase while in $NM_{(-)}$, $m_{1}$ moves oppositely to $m_{2}$ and $m_{3}$. In the third mode, $NM_{(+)}$, $m_{1}$ stays stationary while $m_{2}$ and $m_{3}$ oscillate harmonically exactly out of phase with each other. The analysis performed in Appendix B can be extended to the present case by defining the normal coordinates $X_{0}$, $X_{+}$ and $X_{-}$, which are proportional to $x_{1} + x_{2} + x_{3}$, $x_{2} - x_{3}$ and $x_{1} - x_{2} - x_{3}$, respectively, meaning that any arbitrary motion of the system is a superposition of those three normal modes. The classical dark state is defined according to the EIT-like conditions for each system.

\subsection{EIT in four-level atoms in the inverted-Y configuration}

The effect of two or more electromagnetic fields interacting with multi-level atomic systems has been extensively explored theoretically and experimentally in recent years \cite{Xiao2003}. The absorption spectrum of a variety of four-level atomic systems exposed to three laser fields is characterized by a double dark resonance. This effect is named as double EIT.

\begin{figure}[!ht]
\includegraphics[width=7.5cm]{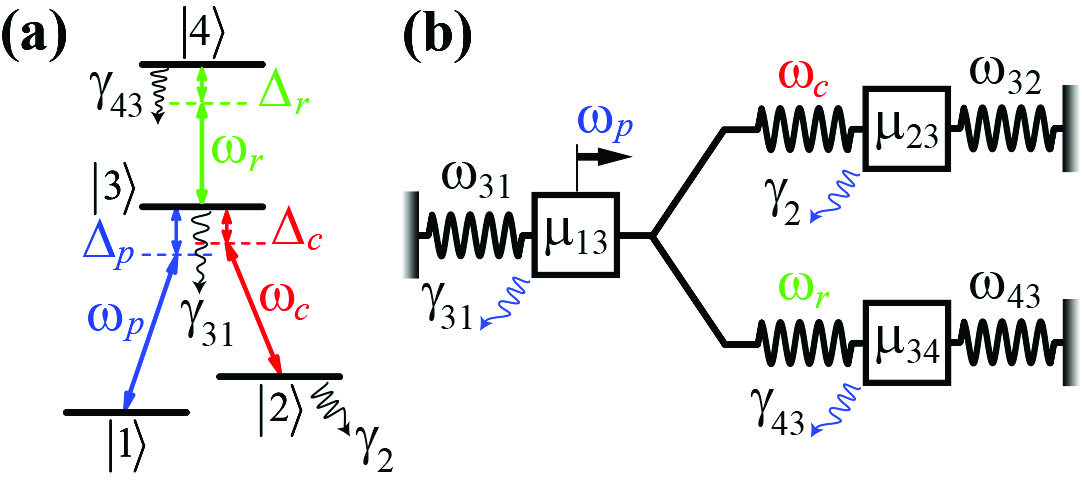}
\caption{(Color online). (a) Schematic energy level diagram of a four-level atom in the inverted-Y configuration, showing three classical electromagnetic fields, probe $(\omega_{p})$, control $(\omega_{c})$ and pump $(\omega_{r})$, coupling the transitions $|1\rangle\leftrightarrow|3\rangle$, $|2\rangle\leftrightarrow |3\rangle$ and $|3\rangle\leftrightarrow|4\rangle$, respectively, and their corresponding detunings. The atomic decay rates are represented by $\gamma_{31}=\Gamma_{31}+\Gamma_{32}+\gamma_{3}$, $\gamma_{43}=\Gamma_{43}+\gamma_{4}$ and $\gamma_{2}$. The classical analog shown in (b) consists of only one force acting on mass $m_{1}$, meaning that $F_{2s}=F_{3s}=0$ in Fig.\ref{TMHCO}.} \label{Yinvertido}
\end{figure}

The four-level atom in the inverted-Y configuration can be seen as a three-level atom in $\Lambda$ configuration, composed by the states $\left\vert 1\right\rangle $, $\left\vert 2\right\rangle $ and $\left\vert 3\right\rangle $, plus a second excited state $\left\vert 4\right\rangle $, as shown in Fig.\ref{Yinvertido}(a). Transitions $\left\vert 1\right\rangle
\leftrightarrow\left\vert 3\right\rangle $ and $\left\vert 2\right\rangle\leftrightarrow\left\vert 3\right\rangle$ interact with the probe and control fields as in the usual three-level $\Lambda$ type. A third coupling field of frequency $\omega_{r}$ and Rabi frequency $2\Omega_{r}$, named as pumping field, couples the transition $\left\vert 3\right\rangle \leftrightarrow \left\vert 4\right\rangle$.

By introducing the dipole and rotating-wave approximations, the time-independent Hamiltonian for this system can be written as

\begin{align} \label{HYinv1}
H  &= -\Delta_{p}\sigma_{11}-\Delta_{c}\sigma_{22}-\Delta_{r}\sigma _{44}-\Omega_{p}\left(  \sigma_{13}+\sigma_{31}\right) \nonumber\\
&- \Omega_{c}\left(  \sigma_{23}+\sigma_{32}\right)  -\Omega_{r}\left(\sigma_{43}+\sigma_{34}\right),
\end{align}
where the detunings are given by $\Delta_{p}=\omega_{31}-\omega_{p}$, $\Delta_{c}=\omega_{32}-\omega_{c}$ and $\Delta_{r}=\omega_{43}-\omega_{r}$. Its dynamics is obtained numerically by solving the master equation for the atomic density operator
\begin{align}
\dot{\rho}  &= - i[H,\rho]+\sum\limits_{m=1,2}\Gamma_{3m}(2\sigma_{m3}\rho\sigma_{3m}-\sigma_{33}\rho-\rho\sigma_{33})\nonumber\\
&+ \Gamma_{43}(2\sigma_{34}\rho\sigma_{43}-\sigma_{44}\rho-\rho\sigma_{44})\nonumber\\
&+ \sum\limits_{n=2,3,4}\gamma_{n}(2\sigma_{nn}\rho\sigma_{nn}-\sigma_{nn}\rho-\rho\sigma_{nn}), \label{e18}
\end{align}
with the polarization decay rate $\Gamma_{43}$ and non-radiative atomic dephasing rate $\gamma_{4}$, accounting for the additional state $\left\vert 4\right\rangle $.

The information about absorption and dispersion of the probe field in the four-level atomic medium is obtained through the reduced electric susceptibility $\tilde{\chi}_{e}(\omega_{p})=\rho_{31}(\omega_{p})$, in analogy with previous definitions. For the inverted-Y system we also used the weak probe field approximation, $\Omega_{p}<<\left(\Omega_{c},\Omega_{r}\right)$, implying that almost all the atomic population is in the ground state $\rho_{11}\approx1$. From the full density-matrix equations of motion and assuming that the values of $\rho_{43}$ and $\rho_{23}$ are approximately zero \cite{Xiao2003}, we solved for the steady state of $\rho$ to find
\begin{equation}
\rho_{31}(\omega_{p})=\frac{\Omega_{p}\left(\delta_{2}-i\gamma_{2}\right)\left(\delta_{4}-i\gamma_{43}\right)}{\Upsilon_{Q}-\Omega_{c}^{2}\left(\delta_{4}-i\gamma_{43}\right) - \Omega_{r}^{2}\left(\delta_{2}-i\gamma_{2}\right)}, \label{RhoY}
\end{equation}
where $\Upsilon_{Q}=\left(\Delta_{p}-i\gamma_{31}\right) \left(\delta_{2}-i\gamma_{2}\right) \left(\delta_{4}-i\gamma_{43}\right)$, $\gamma_{31}=\Gamma_{31}+\Gamma_{32}+\gamma_{3}$ and $\gamma_{43}=\Gamma_{43}+\gamma_{4}$. Here we introduced the two-photon detunings $\delta_{2}=\Delta_{p}-\Delta_{c}$ and $\delta_{4}=\Delta_{p}-\Delta_{r}$. Note that when $\Omega_{r}=0$, eq.\eqref{RhoY} reduces to eq.\eqref{e5} for the three-level EIT-$\Lambda$ configuration.

The classical analog to demonstrate double EIT in four-level atoms in the inverted-Y configuration was proposed by Serna et al. \cite{Serna2011}. They used a mechanical system comprised by three coupled harmonic oscillators and also an electric analog composed by three coupled RLC circuits. Here we used the same configuration as in \cite{Serna2011} in order to identify an one-to-one correspondence between the classical and quantum dynamic variables for this system.

Its corresponding reduced mechanical susceptibility $\tilde{\chi}_{M}(\omega_{s})=\rho_{co}(\omega_{s})$ is obtained from eqs.\eqref{CETL} by setting $F_{2s}=F_{3s}=0$ and solving for the displacement of particle 1 for $\phi_{1}=0$,
\begin{equation}
\rho_{co}(\omega_{s})=\frac{\Omega_{s}\left(\Delta_{2}-i\gamma_{2}\right)\left(\Delta_{3}-i\gamma_{3}\right)}{\Upsilon_{C}-\Omega_{12}^{2}\left(\Delta_{3}-i\gamma_{3}\right)  -\Omega_{13}^{2}\left(\Delta_{2} - i\gamma_{2}\right)}, \label{e19}
\end{equation}
where $\Upsilon_{C}=\left(\Delta_{1}-i\gamma_{1}\right) \left(\Delta_{2}-i\gamma_{2}\right)\left(\Delta_{3}-i\gamma_{3}\right)$, the coupling rates $\Omega_{12}=\omega_{12}^{2}/2\sqrt{\omega_{1}\omega_{2}}$, $\Omega_{13}=\omega_{13}^{2}/2\sqrt{\omega_{1}\omega_{3}}$ and the pumping rate $\Omega_{s}=\sqrt{F_{1}^{2}/2m\omega_{1}}$. As we have discussed in Sec.IIA the coupling-field detunings $\Delta_{c}$ and $\Delta_{r}$ in eq.\eqref{RhoY} can be reproduced readily in the classical system by setting $\Delta_{1}=\Delta_{s}$, $\Delta_{2}=\Delta_{s}-\Delta_{21}$ and $\Delta_{3}=\Delta_{s}-\Delta_{31}$, where $\Delta_{21}$ and $\Delta_{31}$ account for the detuning between the frequencies of the oscillators 2-1 and 3-1, respectively. For perfect resonances $\Delta_{c}=\Delta_{r}=0$, the classical detunings are reduced to $\Delta_{1}=\Delta_{2}=\Delta_{3}=\Delta_{s}$. Note that even for $k_{2}=k_{3}$ we have $\omega_{2}\neq\omega_{3}$ so that, for the resonance case the analog is complete by adjusting the detunings to be identical through $k_{1}$, $k_{12}$ and $k_{13}$.

Comparing $\rho_{31}(\omega_{p})$, eq.\eqref{RhoY}, and $\rho_{co}(\omega_{s})$, eq.\eqref{e19}, we identify classically each parameter of the atomic system as in Table \ref{AYinv}. The classical analog is illustrated in Fig.\ref{Yinvertido}(b). As shown before, each atomic dipole-allowed transition corresponds to a harmonic oscillator in the mechanical system. Then, the addition of state $\left\vert 4\right\rangle $ and the coupling field of frequency $\omega_{r}$ imply the addition of one more harmonic oscillator ($m_{3}$), to account for the atomic transition $|3\rangle\leftrightarrow|4\rangle$, and a second coupling spring ($k_{13}$) to communicate energy to the pumped oscillator $m_{1}$.

\begin{table} \label{AYinv}
\caption{Classical analog of EIT-like for the four-level atom in an inverted-Y configuration (EIT-4Y) using three mechanical coupled harmonic oscillators (3-MCHO).}
\begin{center}
\begin{tabular} {c c}
\hline\hline
EIT-4Y $\left(\rho_{31}\right)$ & 3-MCHO $\left(\rho_{co}\right)$ \\[1ex]
\hline
$\Delta_{p}$ & $\Delta_{1}$\\
$\delta_{2}$ & $\Delta_{2}$\\
$\delta_{4}$ & $\Delta_{3}$\\
$\Omega_{p}$ & $\Omega_{s}$\\
$\Omega_{c}$ & $\Omega_{12}$\\
$\Omega_{r}$ & $\Omega_{13}$\\
$\gamma_{31}$ & $\gamma_{1}$\\
$\gamma_{2}$ & $\gamma_{2}$\\
$\gamma_{43}$ & $\gamma_{3}$\\[1ex]
\hline
\end{tabular}
\end{center}
\end{table}

\begin{figure}[!ht]
\includegraphics[width=8.5cm]{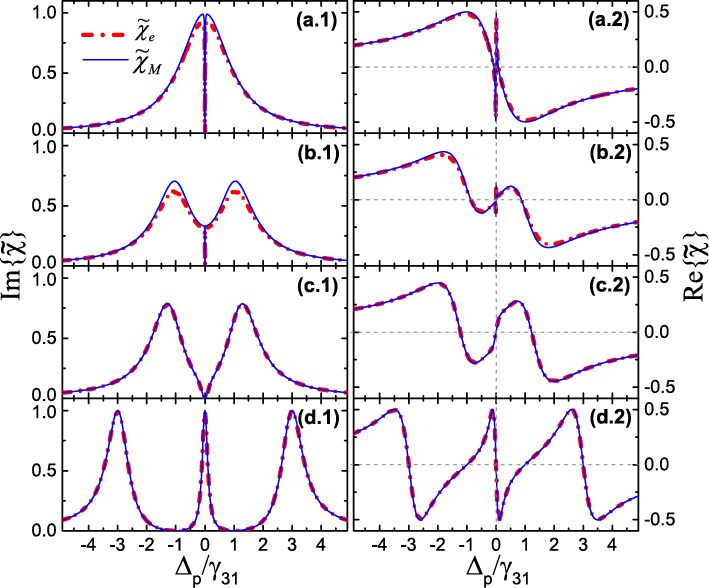}
\caption{(Color online). Imaginary and real parts of the reduced electric susceptibility ($\tilde{\chi}_{e}$) vs normalized probe-atom detuning $\Delta_{p}/\gamma_{31}$ for the four-level atom in a inverted-Y configuration in comparison with its classical counterpart ($\tilde{\chi}_{M}$) obtained using three coupled harmonic oscillators. The parameters are $\Omega
_{p}=0.02\gamma_{31}$, $\Gamma_{43}=0.5\gamma_{31}$, $\gamma_{2}=0.0$, (a) $\Omega_{c}=\Omega_{r}=0.08\gamma_{31}$, (b) $\Omega_{c}=0.08\gamma_{31}$, $\Omega_{r}=1.0\gamma_{31}$, (c) $\Omega_{c}=0.8\gamma_{31}$, $\Omega_{r}=1.0\gamma_{31}$ and (d) $\Omega_{c}=\Omega_{r}=2.0\gamma_{31}$. The coupling-field detunings $\Delta_{c}$, $\Delta_{r}$ are zero in (a), (b), (c)
and (d) $\Delta_{c}=1.0\gamma_{31}$, $\Delta_{r}=-1.0\gamma_{31}$. For the classical system we use the same set of parameters following the analog presented in table \ref{AYinv}.} \label{YinvA}
\end{figure}

The imaginary and real parts of the reduced electric susceptibility $\tilde{\chi}_{e}(\omega_{p})$ are depicted in Fig.\ref{YinvA} as a function of the normalized probe-atom detuning $\Delta_{p}/\gamma_{31}$ in comparison with its classical counterpart $\tilde{\chi}_{M}(\omega_{s})$. Figures \ref{YinvA}(a) and \ref{YinvA}(b) show disagreement between the results,
meaning that the condition $\Omega_{p}<<\left(  \Omega_{c},\Omega_{r}\right)$ is not deeply satisfied and part of the atomic population is not in the ground state $\left\vert 1\right\rangle $. In Fig.\ref{YinvA}(c) and Fig.\ref{YinvA}(d) the condition is satisfied with classical and quantum results showing excellent agreement. The classical dark state in this case is also produced when oscillator 1 stays stationary while oscillators 2 and 3 oscillate harmonically. Note that when $\omega_{s} = \omega_{1} = \sqrt{\left(k_{1}+k_{12}+k_{13}\right)/m}$ the system is pumped in the range between the normal frequencies $\omega_{0}$ and $\omega_{-}$, which is a region of high probability to occur interference between the normal modes $NM_{(0)}$ and $NM_{(-)}$. Once $x_{1} = 0$, it is featured by zero absorption power of oscillator 1, which is equivalent to $\tilde{\chi}_{M} = 0$ for zero detuning.

Figure \ref{YinvA}(d) shows that a third resonance peak appears as a consequence of making the coupling-atom detunings $\Delta_{c}$ and $\Delta_{r}$ different of zero. If we set $\Omega_{c} = \Omega_{r}$ the peaks become symmetric giving rise to two transmission windows, which characterizes double EIT. By manipulating the parameters of the system we can control the two EIT dips from a narrow to a wider splitting of the Autler-Townes doublets. We see that all these resonant features can be reproduced with the mechanism of classical interference of the normal modes of the three coupled harmonic oscillators in the displacement of oscillator 1.

\subsection{EIT in four-level atom in a tripod configuration}

The four-level atom in a tripod configuration is also based on a three-level EIT system and it is promising for many applications, ranging from the realization of polarization quantum phase gates to quantum information processes \cite{Friedmann2004, Corbalan2004, Malakyan2004, Wang2008}.

\begin{figure}[!ht]
\includegraphics[width=8.5cm]{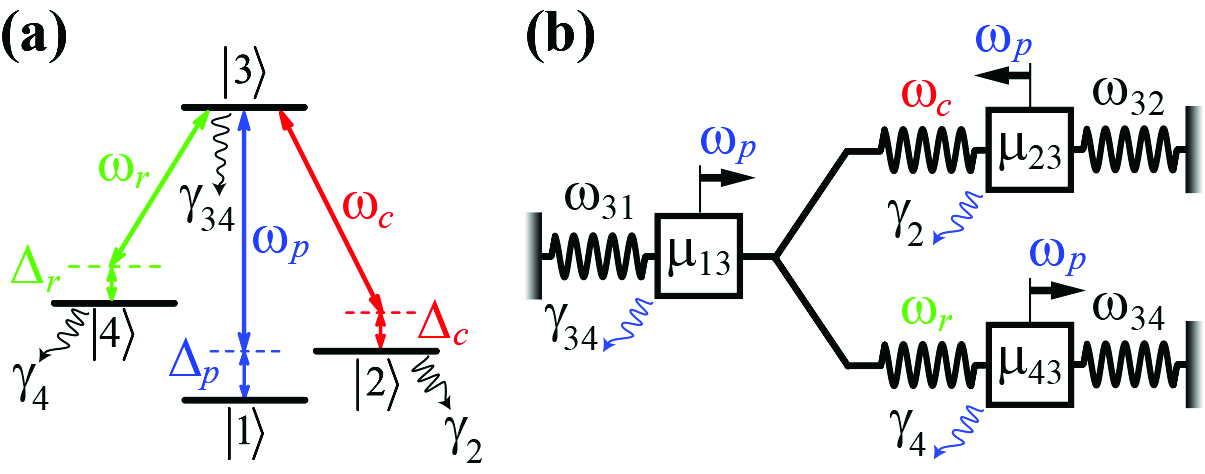}
\caption{(Color online). (a) Schematic energy level diagram of a four-level atom in a tripod configuration, showing three classical electromagnetic fields, probe $(\omega_{p})$, control $(\omega_{c})$ and pump $(\omega_{r})$, coupling the transitions $|1\rangle\leftrightarrow|3\rangle$, $|2\rangle\leftrightarrow |3\rangle$ and $|3\rangle\leftrightarrow|4\rangle$, respectively, and their corresponding detunings. The atomic decay rates are represented by $\gamma_{34}=\Gamma_{31}+\Gamma_{32}+\Gamma_{34}+\gamma_{3}$, $\gamma_{2}$ and $\gamma_{4}$. The classical analog is obtained considering a force acting in each harmonic oscillator with phases $\phi_{1}=\phi_{3}=0$ and $\phi_{2}=\pi$, as shown in (b).} \label{Tripod}
\end{figure}

Differently of the inverted-Y configuration, here the atomic level $\left\vert 4\right\rangle $ is a ground state, see Fig.\ref{Tripod}(a). The time-independent Hamiltonian is essentially the same as eq.\eqref{HYinv1} and the master equation is slightly modified as,
\begin{align}
\dot{\rho} &= -i[H,\rho]+\sum\limits_{m=1,2,4}\Gamma_{3m}(2\sigma_{m3}\rho\sigma_{3m}-\sigma_{33}\rho-\rho\sigma_{33})\nonumber\\
&+ \sum\limits_{n=2,3,4}\gamma_{n}(2\sigma_{nn}\rho\sigma_{nn}-\sigma_{nn}\rho-\rho\sigma_{nn}),\label{e20}
\end{align}
where we introduce the polarization decay rate $\Gamma_{34}$ of the excited level $|3\rangle$ to the level $|4\rangle$.

In the same way as in the inverted-Y configuration the response of the probe field is given by the reduced electric susceptibility $\tilde{\chi}_{e}=\rho_{31}$. Solving for $\rho_{31}$ and considering the limit of low atomic excitation $\rho_{11}\approx1$ we have,
\begin{equation}
\rho_{31}=\frac{\Omega_{p}\left(\Delta_{p}-i\gamma_{2}\right) \left(\Delta_{p}-i\gamma_{4}\right) - \Omega_{p}\Omega_{c}\Upsilon_{23}-\Omega_{p}\Omega_{r}\Upsilon_{43}}{\Upsilon_{Q}-\Omega_{c}^{2}\left(\Delta_{p}-i\gamma_{4}\right) - \Omega_{r}^{2}\left(\Delta_{p}-i\gamma_{2}\right)}, \label{FQtripod}
\end{equation}
where $\Upsilon_{23}=\left(\Delta_{p}-i\gamma_{4}\right) \rho_{23}$, $\Upsilon_{43}=\left(\Delta_{p}-i\gamma_{2}\right) \rho_{43}$ and $\Upsilon_{Q}=\left(\Delta_{p}-i\gamma_{34}\right) \left(\Delta_{p}-i\gamma_{2}\right) \left(\Delta_{p}-i\gamma_{4}\right)$ with $\gamma_{34}=\Gamma_{31}+\Gamma_{32}+\Gamma_{34}+\gamma_{3}$.

\begin{figure}[!ht]
\includegraphics[width=7cm]{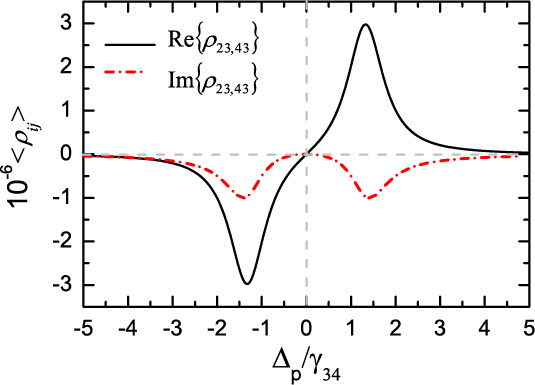}
\caption{(Color online). Imaginary and real parts of $\rho_{23}$ and $\rho_{43}$ vs the normalized probe-atom detuning $\Delta_{p}/\gamma_{34}$ for perfect atom-field resonances $\Delta_{c}=\Delta_{r}=0$, using the parameters $\Omega_{p}=0.002\gamma_{34}$, $\Omega_{c}=\Omega_{r}=1.0\gamma_{34}$ and $\gamma_{2}=\gamma_{4}=0$.} \label{sigma3234}
\end{figure}

The real and imaginary parts of the nondiagonal density matrix element $\rho_{23}$ are identical to the same for $\rho_{43}$, as shown in Fig.\ref{sigma3234}. Despite their small values they are not neglected here, like in the inverted-Y configuration. Note that the real parts of $\rho_{23,43}$ change their signal with $\Delta_{p}$, while the signal of the
imaginary parts are kept the same. These details are essential to obtain the correct classical analog for the atomic tripod configuration.

If we consider $\Omega_{r}=0$ in eq.\eqref{FQtripod} we end up with,
\begin{equation}
\rho_{31}=\frac{\Omega_{p}\left(\Delta_{p}-i\gamma_{2}\right) - \Omega_{p}\Omega_{c}\rho_{23}}{\left(\Delta_{p}-i\gamma_{34}\right) \left(\Delta_{p}-i\gamma_{2}\right) - \Omega_{c}^{2}}. \label{FQtripodB}
\end{equation}

Apart from the dimensionless term $\rho_{23}$, the equation above has the same form of a mechanical model comprised by two harmonic oscillators with two forces acting on particles 1 and 2 out of phase by $\pi$. In eqs.\eqref{CETL} we would have $F_{2}=-F_{1}$ for $k_{13}=0$, or $F_{3}=-F_{1}$ for $k_{12}=0$, once the same is observed for $\Omega_{c}=0$. Then, as a first suggestion, one could propose the classical analog for the atomic tripod configuration by considering the forces $F_{2s}$ and $F_{3s}$ out of phase with $F_{1s}$ by $\pi$ , i.e., $\phi_{1}=0$ and $\phi_{2}=\phi_{3}=\pi$. But Fig.\ref{sigma3234} shows that the real parts of $\rho_{23,43}$ are in phase with their corresponding imaginary parts for $\Delta_{p}<0$ and out of phase by $\pi$ for $\Delta_{p}>0$. As additional transitions, $\rho_{23}$ and $\rho_{43}$, represent additional harmonic oscillators we reproduce this effect by assuming only the force acting on particle 2 out of phase by $\pi$ with the force applied on particle 1, meaning that $F_{2}=-F_{1}$ and $F_{3}=F_{1}$. This classical model mimics the EIT features presented by the tripod configuration in very good agreement.

Taking into account the considerations above the reduced mechanical susceptibility is obtained from equations \eqref{CETL} for the displacement of oscillator 1 as follows,
\begin{equation}
\rho_{co}=\frac{\Omega_{s}^{(1)}\left(\Delta_{2}-i\gamma_{2}\right) \left(\Delta_{3}-i\gamma_{3}\right)  -\Omega_{s}^{(2)}\Omega_{12}\Upsilon_{3}+\Omega_{s}^{(3)}\Omega_{13}\Upsilon_{2}}{\Upsilon_{C}-\Omega_{12}^{2}\left(\Delta_{3}-i\gamma_{3}\right) - \Omega_{13}^{2}\left( \Delta_{2}-i\gamma_{2}\right)}, \label{Ctripod}
\end{equation}
where $\Upsilon_{3}=\Delta_{3}-i\gamma_{3}$, $\Upsilon_{2}=\Delta_{2} - i\gamma_{2}$ and $\Upsilon_{C}=\left(\Delta_{1}-i\gamma_{1}\right)\left(\Delta_{2}-i\gamma_{2}\right) \left( \Delta_{3}-i\gamma_{3}\right)$, $\Omega_{12}=\omega_{12}^{2}/2\sqrt{\omega_{1}\omega_{2}}$ and $\Omega_{13}=\omega_{13}^{2}/2\sqrt{\omega_{1}\omega_{3}}$. The mechanical pumping rates are given by $\Omega_{s}^{(j)}=\sqrt{F_{j}^{2}/(2m\omega_{j})}$ and they are related to the force $F_{j}$ acting on the $j$-th oscillator, $j=1,2,3$.

Once there is only one probe field applied to the atomic system with Rabi frequency $\Omega_{p}$, eq.\eqref{FQtripod}, the classical pumping rates have to be the same, i.e., $\Omega_{s}^{(j)}=\Omega_{s}$. Consequently $\omega_{1}=\omega_{2}=\omega_{3}$, implying that $k_{2}=k_{1}+k_{13}$ and $k_{3}=k_{1}+k_{12}$. This also conducts to $\Delta_{1}=\Delta_{2}=\Delta_{3}=\Delta_{s}$. Considering all these conditions, eq.\eqref{Ctripod} becomes identical to eq.\eqref{FQtripod} for the atomic system. The classical analog for each parameter is depicted in table \ref{TableTripod} and illustrated in Fig.\ref{Tripod}(b).

Huang et al. \cite{Huang2013} proposed recently a classical analog for the atomic tripod configuration, considering $F_{1}=0$ and $F_{2}=F_{3}$ in eqs.\eqref{CETL}. According to them their classical analog, or in our terms, their reduced mechanical susceptibility $\tilde{\chi}_{M}^{H}=\rho_{co}^{H}$ is obtained solving for the displacement of oscillators 2 or 3. Using these conditions and the same definitions above we have,
\begin{equation}
\rho_{co}^{H}=\frac{\Omega_{s}\left(\Delta_{s}-i\gamma_{1}\right) \left(\Delta_{s}-i\gamma_{3}\right) - \Omega_{s}\Omega_{13}^{2}+\Omega_{s}\Omega_{12}\Omega_{13}}{\Upsilon_{C}-\Omega_{12}^{2}\left(  \Delta_{s} - i\gamma_{3}\right) - \Omega_{13}^{2}\left(\Delta_{s}-i\gamma_{2}\right)}. \label{RhoCOHuang}
\end{equation}

Comparing eq.\eqref{RhoCOHuang} with $\rho_{31}$, eq.\eqref{FQtripod}, we see that it is not possible to establish a one-to-one classical correspondence for the quantum variables $\Upsilon_{23}=\left(\Delta_{p}-i\gamma_{4}\right)  \rho_{23}$ and $\Upsilon_{43} = \left(\Delta_{p}-i\gamma_{2}\right)\rho_{43}$. According to eq.\eqref{RhoCOHuang} we would have
$\Omega_{c}\Upsilon_{23}\equiv\Omega_{13}^{2}$ and $-\Omega_{r}\Upsilon_{43}\equiv\Omega_{12}\Omega_{13}$. The classical analog for the other variables are shown in table \ref{TableTripod}. Note that we have two constraints for the classical variables in this case, $\gamma_{1}=\gamma_{2}$ and $\Omega_{12}=\Omega_{13}$.

\begin{table}
\caption{Classical analog of EIT-like in a four-level atom in a tripod configuration (EIT-Tripod) using three mechanical coupled harmonic oscillators considering the forces acting on the three particles as $F_{2}=-F_{1}$ and $F_{3}=F_{1}$ for our model (3CO) and $F_{2}=F_{3}$, $F_{1}=0$ for Huang's model (3CO-H) \cite{Huang2013}.}
\begin{center}
\begin{tabular}{c c c }
\hline\hline
EIT-Tripod $\left(\rho_{31}\right)$ & 3CO $\left(\rho_{co}\right)$ & 3CO-H $\left(\rho^{H}_{co}\right)$ \\[1ex]
\hline
$\Delta_{p}$ & $\Delta_{s}$ & $\Delta_{s}$\\
$\Omega_{p}$ & $\Omega_{s}$ & $\Omega_{s}$\\
$\Omega_{c}$ & $\Omega_{12}$ & $\Omega_{12}$, $\Omega_{13}$\\
$\Omega_{r}$ & $\Omega_{13}$ & $\Omega_{12}$, $\Omega_{13}$\\
$\gamma_{34}$ & $\gamma_{1}$ & $\gamma_{1}$\\
$\gamma_{2}$ & $\gamma_{2}$ & $\gamma_{1}, \gamma_{2}$\\
$\gamma_{4}$ & $\gamma_{3}$ & $\gamma_{3}$\\
$\Upsilon_{23}$ & $\Upsilon_{3}$ & -\\
$-\Upsilon_{43}$ & $\Upsilon_{2}$ & -\\[1ex]
\hline
\end{tabular}
\end{center}
\label{TableTripod}
\end{table}

\begin{figure}[!ht]
\includegraphics[width=8.5cm]{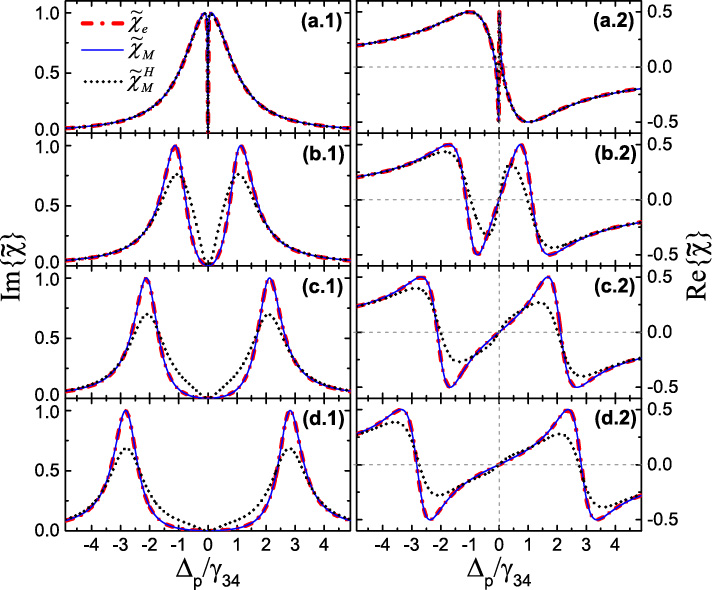}
\caption{(Color online). Imaginary and real parts of the reduced electric susceptibility $\tilde{\chi}_{e}$ vs normalized probe-atom detuning $\Delta_{p}/\gamma_{34}$ for the four-level atom in a tripod configuration in comparison with its classical counterparts $\tilde{\chi}_{M}$, eq.\eqref{Ctripod}, and $\tilde{\chi}_{M}^{H}$, eq.\eqref{RhoCOHuang}, obtained using three coupled harmonic oscillators. The parameters are $\Omega_{p}=0.002\gamma_{34}$, $\Delta_{c}=\Delta_{r}=0$, $\gamma_{2}=\gamma_{4}=0$ for different values of the Rabi frequencies of the coupling $\Omega_{c}$ and pumping $\Omega_{r}$ fields. It is considered $\Omega_{c}=\Omega_{r}$ with values (a) $0.08\gamma_{34}$, (b) $0.8\gamma_{34}$, (c) $1.5\gamma_{34}$ and (d) $2.0\gamma_{34}$. For the classical models we obtain $\tilde{\chi}_{M}$ and $\tilde{\chi}_{M}^{H}$ using the same set of parameters following the analog presented in table \ref{TableTripod}.} \label{Tripod3M}
\end{figure}

In Fig.\ref{Tripod3M} we plot the real and imaginary parts of the reduced electric susceptibility $\tilde{\chi}_{e}$ for the atomic system as a function of the normalized probe-atom detuning $\Delta_{p}/\gamma_{34}$ in comparison with its two classical counterparts $\tilde{\chi}_{M}$ and $\tilde{\chi}_{M}^{H}$ obtained from eqs.\eqref{Ctripod} and \eqref{RhoCOHuang}, respectively. We consider the weak-probe limit $\Omega_{p}<<(\Omega_{c},\Omega_{r})$ with $\Omega_{p}=0.002\gamma_{34}$ for perfect coupling-field resonances $\Delta_{c}=\Delta_{r}=0$ and $\gamma_{2}=\gamma_{4}=0$. For all cases we consider $\Omega_{c}=\Omega_{r}$ owing the constraint obtained from eq.\eqref{RhoCOHuang}, where $\Omega_{12}=\Omega_{13}$.

Figure \ref{Tripod3M}(a) shows that both classical analogs reproduce the EIT features calculated for the atomic tripod system in very good agreement. When the Rabi frequencies of the coupling $(\Omega_{c})$ and pumping $(\Omega_{r})$ fields increase, Figs.\ref{Tripod3M}(b), \ref{Tripod3M}(c) and \ref{Tripod3M}(d), show that only the mechanical susceptibility $\tilde{\chi}_{M}$, given by eq.\eqref{Ctripod}, reproduces satisfactorily the behavior of the atomic system.

Although the impossibility of obtaining a one-to-one correspondence between classical and quantum variables, the classical analog proposed in ref.\cite{Huang2013}, eq.\eqref{RhoCOHuang}, exhibits a similar behavior as the tripod configuration, but total agreement is observed only for small values of $\Omega_{12}$, $\Omega_{13}$. If the EIT-like condition $\Omega_{p}<<(\Omega_{c},\Omega_{r})$ is deeply satisfied, the analog proposed here shows perfect agreement for any set of parameters.

\subsection{Cavity EIT (CEIT)}

In Sec.IIC we have shown the classical analog for a system consisting of a single two-level atom coupled to a single cavity mode. In this section we present for the first time the analog for the extended system considering a three-level atom placed inside an optical cavity. This system also exhibits EIT features being usually referred to as intracavity EIT or simply cavity EIT (CEIT). The optical cavity enhances the main characteristics of EIT, regarding atomic coherence and interference, which may be useful for a variety of fundamental studies and practical applications \cite{Scully1998, Zhu2007, Xiao2008, Souza2013}.

\begin{figure}[!ht]
\includegraphics[width=8.5cm]{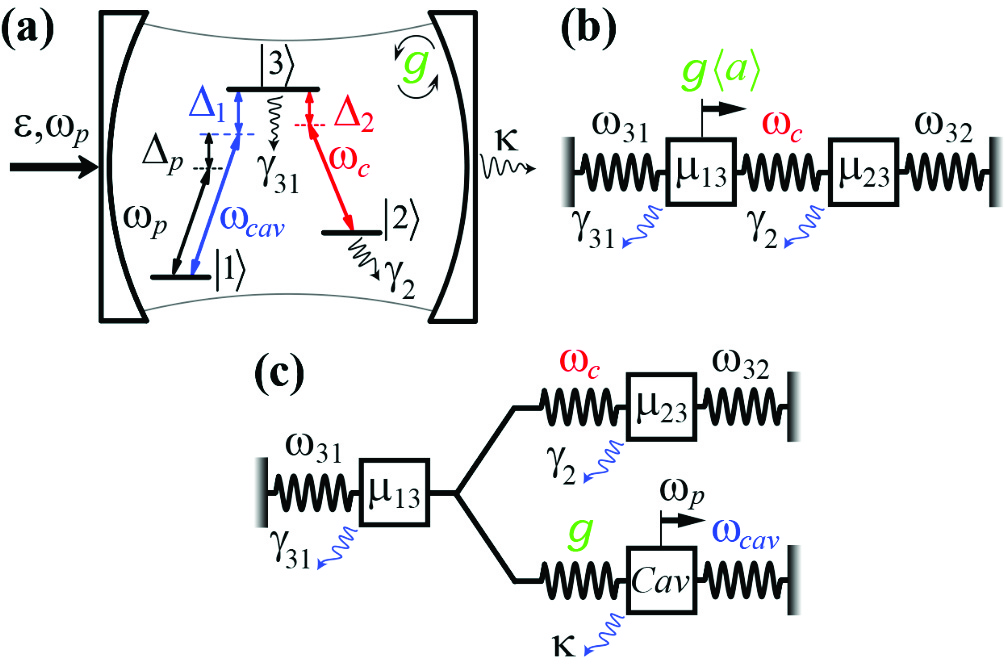}
\caption{(Color online). Three-level atom in a $\Lambda$ configuration inside an optical resonator showing the quantum cavity field with frequency $\omega_{cav}$ and vacuum Rabi frequency $2g$ coupling the atomic transition $|1\rangle\leftrightarrow|3\rangle$. The control field with frequency $\omega_{c}$ couples the transition $|2\rangle\leftrightarrow|3\rangle$ and the probe field with frequency $\omega_{p}$ and strength $\varepsilon$ drives the cavity mode. (b) Classical analog for $\left\langle \sigma_{13}\right\rangle $ given by eq.\eqref{sigma13} corresponding to two coupled harmonic oscillators pumped by the Rabi frequency of the cavity field $g\left\langle a\right\rangle $. (c) Classical analog for each parameter of the CEIT system.} \label{CEIT}
\end{figure}

The system is comprised of a single atom with three energy levels in $\Lambda$ configuration, as in Fig.\ref{EsquemaEIT}(a), coupled to a single electromagnetic mode of frequency $\omega_{cav}$ of an optical resonator, see Fig.\ref{CEIT}(a). The cavity is driven by a coherent field (probe) of strength $\varepsilon$ and frequency $\omega_{p}$. The atomic transitions $|1\rangle\leftrightarrow |3\rangle$ (frequency $\omega_{31}$) and $|2\rangle\leftrightarrow|3\rangle$ (frequency $\omega_{32}$) are coupled by the cavity mode with vacuum Rabi frequency $2g$ and by a classical field (control) with frequency $\omega_{c}$ and Rabi frequency $2\Omega_{c}$, respectively. The time-independent Hamiltonian which describes the atom-field coupling in a rotating frame is given by
\begin{align}
H  &= - \Delta_{p}\sigma_{11}+\left(  \Delta_{1}-\Delta_{2}\right) \sigma_{22} + \Delta_{1}\sigma_{33}+\Delta_{p}a^{\dagger}a \nonumber\\
&+ \left(  ga\sigma_{31}+\Omega_{c}\sigma_{32}+\varepsilon a+h.c.\right), \label{e22}
\end{align}
where the detunings are $\Delta_{p}=\omega_{cav}-\omega_{p}$, $\Delta_{1}=\omega_{31}-\omega_{cav}$ and $\Delta_{2}=\omega_{32}-\omega_{c}$. The master equation for the atom-cavity density operator is the same as eq.\eqref{e14}, where we have to consider the cavity-field decay rate $\kappa$, the polarization decay rates $\Gamma_{3m}$ $(m=1,2)$ of the excited level
$|3\rangle$ to the levels $|m\rangle$ and the non-radiative atomic dephasing rates $\gamma_{n}$ $(n=2,3)$ of states $|n\rangle$.

Similarly to the standard two-level atom-cavity system (CQED), in the EIT-like condition $\Omega_{c} >> g\left\langle a\right\rangle _{max}$, with $\left\langle a\right\rangle _{max} = \varepsilon/\left(\Delta_{p} - i\kappa\right)$, the CEIT system will be limited to the first splitting of the dressed states, Autler-Townes-like effect, separated by $2\sqrt{g^{2} +
\Omega_{c}^{2}}$. Additionally, there are the intracavity dark states which causes an empty-cavity-like transmission, not observed in the two-level CQED configuration. The CEIT dressed states also compose a kind of anharmonic Jaynes-Cummings ladder structure \cite{Souza2013}.

The probe response is given by the reduced atom-cavity susceptibility which is represented by the expectation value of the cavity field $\tilde{\chi}_{CEIT}(\omega_{p})=\left\langle a\right\rangle $. In the steady state $\dot{\rho}=0$ and considering the low atomic excitation limit $\left\langle\sigma_{11}\right\rangle \approx 1$ we have 
\begin{equation}
\left\langle a\right\rangle =\frac{-\varepsilon\left(\delta_{1}-i\gamma_{31}\right) \left(\delta_{2}-i\gamma_{2}\right) + \varepsilon\Omega_{c}^{2}}{\Upsilon_{Q}-\Omega_{c}^{2}\left(  \Delta_{p}-i\kappa\right) - g^{2}\left(\delta_{2}-i\gamma_{2}\right)}, \label{e23}
\end{equation}
where $\Upsilon_{Q}=\left(\delta_{1}-i\gamma_{31}\right) \left(\delta_{2}-i\gamma_{2}\right) \left(\Delta_{p}-i\kappa\right)$ with $\gamma_{31}=\Gamma_{31}+\Gamma_{32}+\gamma_{3}$, $\delta_{1}=\Delta_{p}-\Delta_{1}$ and $\delta_{2}=\delta_{1}-\Delta_{2}$.

Once the atom-cavity system consists of two atomic dipole allowed transitions and one cavity mode, its classical analog is also modeled on three coupled harmonic oscillators. The analysis of the probe response for the tripod system, given by $\rho_{31}$, revealed that more than one mechanical force have to be taken into account in the mechanical configuration. For all other systems considered before we see that the probe field is represented by a coherent force applied only on the harmonic oscillator corresponding to the respective atomic transition or cavity mode.

By inspection of the expectation value of $\sigma_{13}$, written as follows,
\begin{equation}
\left\langle \sigma_{13}\right\rangle =\frac{-g\left\langle a\right\rangle \left(\delta_{2}-i\gamma_{2}\right)}{\left(\delta_{1} - i\gamma_{31}\right)\left(\delta_{2}-i\gamma_{2}\right) - \Omega_{c}^{2}},\label{sigma13}
\end{equation}
we see that, it is basically the equation for two coupled harmonic oscillators pumped by the Rabi frequency of the cavity field $g\left\langle a\right\rangle$, as illustrated in Fig.\ref{CEIT}(b). Thus, for the classical analog of CEIT we also consider only one force applied on the harmonic oscillator representing the cavity mode, which is driven by the probe field.

Then, the classical analog is obtained from eqs.\eqref{CETL} considering $F_{1s}=F_{2s}=0$. Solving for the displacement of particle 3 and considering $\phi_{3}=\pi$ we find for the reduced mechanical susceptibility $\tilde{\chi}_{M}=\rho_{co}$,
\begin{equation}
\rho_{co}(\omega_{s})=\frac{-\Omega_{s}\left(\Delta_{1}-i\gamma_{1}\right)\left(\Delta_{2}-i\gamma_{2}\right) + \Omega_{s}\Omega_{12}^{2}}{\Upsilon_{C}-\Omega_{12}^{2}\left(\Delta_{3}-i\gamma_{3}\right) - \Omega_{13}^{2}\left(\Delta_{2}-i\gamma_{2}\right)}, \label{CCEIT}
\end{equation}
where $\Upsilon_{C}=\left(\Delta_{1}-i\gamma_{1}\right) \left(\Delta_{2}-i\gamma_{2}\right)  \left(\Delta_{3}-i\gamma_{3}\right)$, $\Omega_{12}=\omega_{12}^{2}/2\sqrt{\omega_{1}\omega_{2}}$, $\Omega_{13}=\omega_{13}^{2}/2\sqrt{\omega_{1}\omega_{3}}$ and $\Omega_{s}=\sqrt{F_{3}^{2}/2m\omega_{3}}$. Note that eqs.\eqref{e23} and \eqref{CCEIT} are identical. The classical analog for each parameter of the CEIT system is shown in table \ref{TableCEIT} and illustrated in Fig.\ref{CEIT}(c).

\begin{table}[th]
\caption{Classical analog of EIT-like for the cavity EIT system (CEIT) using three mechanical coupled harmonic oscillators (3-MCHO).}
\begin{center}
\begin{tabular}{c c}
\hline\hline
CEIT $\left(\left\langle a\right\rangle \right)$ & 3-MCHO $\left(\rho_{co}\right)$ \\[1ex]
\hline
$\delta_{1}$ & $\Delta_{1}$\\
$\delta_{2}$ & $\Delta_{2}$\\
$\Delta_{p}$ & $\Delta_{3}$\\
$\varepsilon$ & $\Omega_{s}$\\
$\Omega_{c}$ & $\Omega_{12}$\\
$g$ & $\Omega_{13}$\\
$\gamma_{31}$ & $\gamma_{1}$\\
$\gamma_{2}$ & $\gamma_{2}$\\
$\kappa$ & $\gamma_{3}$\\[1ex]
\hline
\end{tabular}
\end{center}
\label{TableCEIT}
\end{table}

\begin{figure}[!ht]
\includegraphics[width=8.5cm]{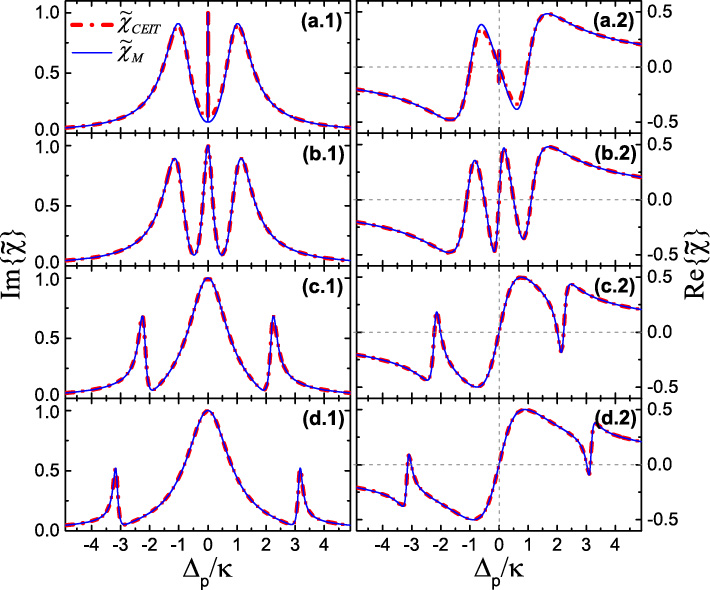}
\caption{(Color online). Imaginary and real parts of the reduced atom-cavity electric susceptibility $\tilde{\chi}_{CEIT}$ vs the normalized probe-cavity detuning $\Delta_{p}/\kappa$ for the CEIT system in comparison with its classical counterpart $\tilde{\chi}_{M}$ for $\Omega_{p}=0.02\kappa$, $g=1.0\kappa$, $\gamma_{31}=0.1\kappa$, $\gamma_{2}=0$, $\Delta_{1}=\Delta_{2}=0$ and different values of the Rabi frequency of the control field (a) $\Omega_{c}=0.02\kappa$, (b) $0.5\kappa$, (c) $2.0\kappa$ and (d) $3.0\kappa$. For the classical system we use the same set of parameters following the analog presented in table \ref{TableCEIT}.} \label{CEITA}
\end{figure}

\begin{figure}[!ht]
\includegraphics[width=8.5cm]{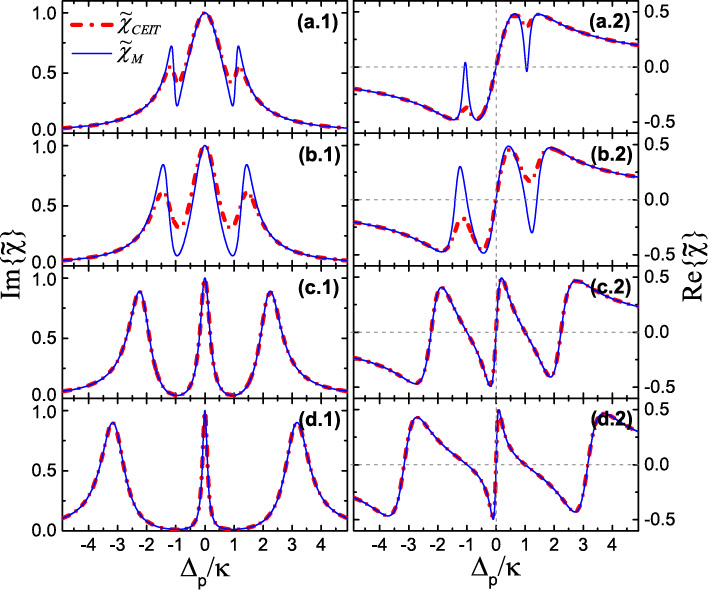}
\caption{(Color online). The same as in Fig.\ref{CEITA} for $\Omega_{c}=1.0\kappa$, $\gamma_{31}=0.1\kappa$, $\gamma_{2}=0$, $\Delta_{1}=\Delta_{2}=0$ and (a) $\Omega_{p}=0.5\kappa$, $g=0.5\kappa$, (b) $0.5\kappa$, $1.0\kappa$, (c) $0.02\kappa$, $2.0\kappa$ and (d) $0.02\kappa$, $3.0\kappa$.} \label{CEITB}
\end{figure}

Figures \ref{CEITA} and \ref{CEITB} show the real and imaginary parts of the reduced atom-cavity susceptibility $\tilde{\chi}_{CEIT}$ vs the normalized probe-cavity detuning $\Delta_{p}/\kappa$ for perfect atom-field resonances $\Delta_{1} = \Delta_{2} = 0$ in comparison with its classical counterpart $\tilde{\chi}_{M}$. The Rabi frequency of the probe field is set to be $\Omega_{p} = 0.02\kappa$ in Fig.\ref{CEITA}, Fig.\ref{CEITB}(c), Fig.\ref{CEITB}(d) and $\Omega_{p} = 0.5\kappa$ in Fig.\ref{CEITB}(a), Fig.\ref{CEITB}(b), while the dissipation rates are fixed at $\gamma_{31} = 0.1\kappa$, $\gamma_{2} = 0$. In Fig.\ref{CEITA} the vacuum Rabi frequency is fixed at $g = 1.0\kappa$ and the steady state of $\left\langle a\right\rangle$ is calculated for different values of the Rabi frequency of the control field $\Omega_{c}$. In Fig.\ref{CEITB} we do the opposite, fixing $\Omega_{c} = 1.0\kappa$ and varying $g$.

Note that there is a small difference between the classical and quantum results in Fig.\ref{CEITA}(a). If we increase the magnitude of $\Omega_{p}$ the difference becomes more pronounced as displayed in Figs.\ref{CEITB}(a) and \ref{CEITB}(b). In these cases the CEIT condition $\Omega_{c}>>g\left\langle a\right\rangle _{max}$ is not deeply satisfied and
$\left\langle \sigma_{11}\right\rangle \neq1$. For all other set of parameters the results show perfect agreement.

The classical dark state, equivalent to the intracavity dark state of the CEIT system, is now observed when oscillator 3 is driven resonantly $\omega_{s} = \omega_{3} = \sqrt{(k_{3} + k_{13})/m}$. Note that this is exactly the resonance frequency $\omega_{+}$ of the normal mode $NM_{(+)}$, where $m_{1}$ stays stationary while $m_{2}$ and $m_{3}$ oscillate harmonically
out of phase with each other. Thus, the classical dark state is naturally identified as a peak in $\omega_{3}$, meaning that the power transferred from the harmonic source to oscillator 3 is total and featured by Im$\left\{\tilde{\chi}_{M}\right\} = 1$ in Fig.\ref{CEITA} and Fig.\ref{CEITB} for zero detuning.

\begin{figure}[!ht]
\includegraphics[width=7cm]{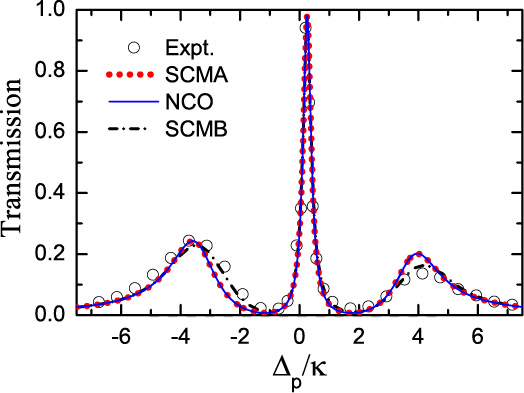}
\caption{(Color online). Experimental transmission spectrum (open circles) vs normalized probe-cavity detuning $\Delta_{p}/\kappa$ for the CEIT system reported in ref.\cite{Muecke2010} for $N\approx15$ atoms in comparison with a semiclassical model and the classical harmonic oscillators. The parameters used for the semiclassical theory, which considers $15$ resting atoms (SCMA - red dotted line), are $\varepsilon=\sqrt{0.02}\kappa$, $g=0.85\kappa$, $\Omega_{c}=1.5\kappa$, $\gamma_{31}=1.04\kappa$, $\gamma_{2}=0.001\kappa$, $\Delta_{1}=-0.3\kappa$, $\Delta_{2}=0$. For the mechanical system, solid blue line (NCO), we make use of the classical analog for $N$ oscillators in eq.\eqref{e25} to calculate $\rho_{Nco}^{\ast}\rho_{Nco}$, using the same set of parameters according to table \ref{TableCEIT} and the analog for the atom-cavity detuning $\Delta_{13}=-0.3\gamma_{3}$. The black dash-dotted line is obtained from the same semiclassical theory as SCMA, but considering the atoms inside the cavity in movement (SCMB). This is performed by changing randomly the parameters $g$, $\Delta_{1}$ and $\Delta_{2}$ in an interval of values specified from experimental considerations.} \label{CEITT}
\end{figure}

Figure \ref{CEITT} displays the transmission spectrum of cavity EIT obtained experimentally by M\"{u}cke \textit{et al.} for 15 atoms, on average, trapped inside a high finesse cavity \cite{Muecke2010}, in comparison with a semiclassical and the classical analog models. As mentioned before, the semiclassical model is obtained from the semiclassical approximation
$\left\langle a\sigma\right\rangle \rightarrow\left\langle a\right\rangle\left\langle \sigma\right\rangle$ where only the field is treated classically. It means that the quantized nature of the three-state atom is respected with $\left\langle a\sigma_{11}\right\rangle \neq\left\langle a\right\rangle $, differently from the full classical case given by eq.\eqref{e23}. The red dotted line in Fig.\ref{CEITT}, named as SCMA, shows the semiclassical result for $N=15$ resting atoms and the black dash-dotted line (SCMB) shows the same semiclassical model but considering atomic motion as in ref.\cite{Muecke2010}. The parameters were adjusted in order to obtain the best fitting. The dephasing rate of state $\left\vert 2\right\rangle$ and the atom-cavity detuning, for example, were set to be $\gamma_{2}=0.001\kappa$ and $\Delta_{1}=-0.3\kappa$, respectively, owing the decreasing in the transmission and the shifting of the central intracavity dark state peak.

We can model mechanically $N$\ atoms by considering $N$\ pairs of harmonic oscillators, like in Fig.\ref{CEIT}(b), coupling independently to oscillator 3, which represents the driven cavity mode. The dynamics of the three-level atom pumped by the Rabi frequency of the cavity can be obtained from the displacement of particle 1 in eqs.\eqref{CETL}. Substituting $N_{2}$\ from eq.\eqref{CETLB} in eq.\eqref{CETLA} we have,
\begin{equation}
N_{1} = \frac{\Omega_{13}\tilde{N}_{3}\left(\Delta_{2}-i\gamma_{2}\right)}{\left(\Delta_{1}-i\gamma_{1}\right)\left(\Delta_{2}-i\gamma_{2}\right)  -\Omega_{12}^{2}}, \label{e24}
\end{equation}
where $\tilde{N}_{3}=\sqrt{\omega_{3}/\omega_{1}}N_{3}$. Note that eq.\eqref{e24} is the classical analog for $\left\langle \sigma_{13}\right\rangle$ given by eq.\eqref{sigma13}. It represents the mechanical atom being pumped by the third harmonic oscillator with pumping rate $\Omega_{13}\tilde{N}_{3}$, in analogy to the Rabi frequency of the cavity field $g\left\langle a\right\rangle$ in the quantum model. Then, if we want to model mechanically $N$ atoms independently coupled to a single cavity mode we have to consider $N\times N_{1}$ in eq.\eqref{CETLC}. Thus, substituting eq.\eqref{e24} in eq.\eqref{CETLC} for $\phi_{3}=\pi$ we end up with,
\begin{equation}
\rho_{Nco}=\frac{-\Omega_{s}\left(\Delta_{1}-i\gamma_{1}\right) \left(\Delta_{2}-i\gamma_{2}\right) + \Omega_{s}\Omega_{12}^{2}}{\Upsilon_{C}-\Omega_{12}^{2}\left(\Delta_{3}-i\gamma_{3}\right) - N\Omega_{13}^{2}\left(\Delta_{2}-i\gamma_{2}\right)}. \label{e25}
\end{equation}

We see that the only difference between eqs.\eqref{CCEIT} and \eqref{e25} is to change the mechanical coupling rate $\Omega_{13}$ for the effective coupling $\Omega_{13}^{(eff)}=\sqrt{N}\Omega_{13}$, where $N$ is the number of pairs of harmonic oscillators as in Fig.\ref{CEIT}(b). Then, to resemble the quantum mechanical average photon number $\left\langle a^{\dagger}a\right\rangle $, which provides the transmission spectrum depicted in Fig.\ref{CEITT}, we have to calculate $\rho_{Nco}^{\ast}\rho_{Nco}$ from eq.\eqref{e25} for $N=15$. As stated before the atom-cavity detuning can be modeled by setting $\Delta_{3}=\Delta_{s}$ and $\Delta_{1}=\Delta_{s}+\Delta_{13}$, where $\Delta_{13}$ accounts for the detuning of the resonant
frequencies between oscillators 1-3.

Using the same set of parameters for the semiclassical model, following the analog depicted in table \ref{TableCEIT}, the full classical result is plotted in Fig.\ref{CEITT}, solid blue line, showing excellent agreement with the semiclassical model SCMA. It indicates that the experiment was performed by considering the CEIT conditions deeply, where $\left\langle \sigma_{11}\right\rangle \approx1$, once the difference between the experimental data and the SCMA theory is solved by taking into account the movement of the atoms inside the cavity, which is corroborated by the SCMB model.

\section{Conclusions}

In this work we showed that mechanical analogs can be obtained for atomic systems which present EIT-related phenomena, if they are considered deeply in the EIT-like conditions. In this case atoms and single cavity modes behave as oscillating dipoles and all dissipative and coherent atom-field processes can be reproduced with systems composed by coupled damped harmonic oscillators. The frequencies of the spectral lines of the atom are equivalent to the natural oscillation frequencies of the oscillators, showing that each atomic-dipole allowed transition corresponds to a classical harmonic damped oscillator. We also showed that the classical dark state is caused by a destructive interference between the normal modes of the system in the displacement of the driven oscillator, and it is observed in analogous conditions with the dark state of the corresponding EIT system.

Through the concept of mechanical susceptibility, with its imaginary part corresponding to the power absorbed by the driven oscillator and its real part related to its amplitude, the classical models presented here describe correctly the action of the atom interacting with an electromagnetic field, reproducing the imaginary and real behavior of the electric susceptibility, respectively. Nevertheless, when the population of the atomic system is shared between its bare states ($\rho_{11} \neq 1$) or when anharmonic effects
takes place, owing the excitation of high energy states, the classical models does not provide a detailed description of the phenomena the way the full quantum theory does. It would be interesting to introduce anharmonicities in the dynamics of the coupled oscillators in order to further explore the connection between these with quantum effects when the EIT-like conditions are not deeply prescribed.

Furthermore, the probe response of driven cavity modes and atom-cavity configurations provide a physical interpretation for the average photon annihilation operator $\left\langle a\right\rangle$, revealing that it can be directly related to the electric susceptibility of the system.

In conclusion, the fact that we can reproduce the phenomenology of EIT with classical harmonic oscillators does not mean EIT is a classical phenomenon. We are just showing that the quantum interference process behind EIT has its equivalent in classical systems, where two or more normal modes interfere to each other to perform such phenomenologies. The patterns of interference observed in the mechanical scheme can be considerably useful to provide a general mapping of EIT-like systems into a variety of classical systems for practical device applications without the necessity of sophisticated technologies required for atomic systems.

\begin{acknowledgments}
We acknowledge fruitful discussions with D. Z. Rossatto. J. A. S. and C.J.V.-B gratefully acknowledge support by the Brazilian founding agency S\~{a}o Paulo Research Foundation (FAPESP) grants \#2013/01182-5, \#2013/04162-5, \#2014/07350-0 and \#2012/00176-9, the Brazilian National Council of Scientific and Technological Development (CNPq) and the Brazilian National Institute of Science and Technology for Quantum Information (INCT-IQ).
\end{acknowledgments}

\section{Appendix}

\subsection {The dynamics of two-coupled harmonic oscillators}

In this appendix we used the Hamiltonian formalism to show that, additionally to the steady-state solution of the EIT system, its dynamics is also equivalent to the dynamics of two coupled harmonic oscillators. Hence, we showed how to obtain $\rho_{co}$, drawn from the Newtonian formalism in Sec.\ref{SecEIT}, eq.\eqref{e9}, using the Hamiltonian of the system.

If we recall from introductory physics the total Hamiltonian for two coupled harmonic oscillators is obtained from the displacement $x_{j}$ and linear momentum $p_{j}$ of the $j$th oscillator as
\begin{eqnarray}\label{HamiltA}
H = \sum^{2}_{j=1}\left(\frac{p_{j}^{2}}{2m} + \frac{1}{2}m\omega_{j}^{2}x_{j}^{2}\right) - m\omega_{12}^{2} x_{1}x_{2} - x_{1}F_{s}(t),
\end{eqnarray}
where we consider the masses to be equal to $m_{1,2} = m$, $\omega_{j}^{2} = \left(k_{j} + k_{12}\right)/m$ ($j = 1,2$), $\omega_{12}^{2} = k_{12}/m$ and the force applied on oscillator 1, $F_{s}(t) = Fe^{-i(\omega_{s} + \phi_{s})t} + c.c.$ for $\phi_{s} = 0$, as illustrated in Fig.\ref{EsquemaEIT}(b). By defining the classical variables $\alpha = \left(m\omega_{1}x_{1} + ip_{1}\right)/\sqrt{2\hbar m\omega_{1}}$ and $\beta = \left(m\omega_{2}x_{2} + ip_{2}\right)/\sqrt{2\hbar m\omega_{2}}$ and considering the simplified case where the natural frequencies of the oscillators are the same, $\omega_{1,2} = \omega$ meaning that $k_{1,2} = k$, the equation above for $\hbar = 1$ takes the form,
\begin{eqnarray}\label{DefOs}
H &=& \omega\left(\alpha^{*}\alpha + \beta^{*}\beta\right) - \frac{\omega_{12}^{2}}{2\omega}\left(\alpha^{*}\beta^{*} + \alpha\beta + \alpha^{*}\beta + \alpha\beta^{*}\right)
\nonumber\\
&-& \sqrt{\frac{F^{2}}{2m\omega}}\left(\alpha^{*} + \alpha\right)\left(e^{i\omega_{s}t} + e^{-i\omega_{s}t}\right).
\end{eqnarray}

The same way as in eq.\eqref{e7} the coupling rate between particles $1$ and $2$ is defined as $\Omega_{12} = \omega_{12}^{2}/2\omega$. Here we are able to find a direct expression for the pumping rate $\Omega_{s}$ as a function of the parameters of the classical system without the necessity of considering the constant $C_{1}$, like in eq.\eqref{e8}. From eq.\eqref{DefOs} we have $\Omega_{s} = \sqrt{F^{2}/2m\omega}$, which is analogous to the Rabi frequency of the probe field ($\Omega_{p}$). 

Now we make an approximation in order to discard fast oscillatory terms like $e^{\pm 2i\omega_{s}t}$ for $\omega \approx \omega_{s}$. This is similar to the rotating wave approximation used in the quantum case. By performing the transformation $\alpha(t) = \tilde{\alpha}(t) e^{-i\omega t}$, likewise for $\beta$, we have,
\begin{eqnarray}\label{HamiA}
H &=& \omega\left(\alpha^{*}\alpha + \beta^{*}\beta\right) - \Omega_{12}\left(\alpha^{*}\beta + \alpha\beta^{*}\right) \nonumber\\
&-& \Omega_{s}\left(\alpha e^{i\omega_{s}t} + \alpha^{*} e^{-i\omega_{s}t}\right).
\end{eqnarray}

From the Poisson brackets $\dot{\rho} = \left\{\rho,H\right\} = -i\partial H/\partial \rho^{*}$ $(\rho = \alpha, \beta)$ the time evolution of $\alpha$ and $\beta$ are given by,
\begin{subequations}\label{HRhos}
\begin{align}
\dot{\alpha} &= -i\left( \omega\alpha - \Omega_{12}\beta - \Omega_{s}e^{-i\omega_{s}t} -i\gamma_{1}\alpha\right),\\
\nonumber\\
\dot{\beta} &= -i\left( \omega\beta - \Omega_{12}\alpha -i\gamma_{2}\beta \right),
\end{align}
\end{subequations}
where we have added phenomenologically the dissipation terms $\gamma_{1}$ and $\gamma_{2}$ in analogy to the master equation formalism. By performing the transformation $\alpha(t) = \rho_{\alpha}(t) e^{-i\omega_{s}t}$, the same way for $\beta$, eqs.\eqref{HRhos} are writen as
\begin{subequations}\label{HRhosB}
\begin{align}
\dot{\rho}_{\alpha} &= -i \left\{\left( \Delta_{s} - i\gamma_{1} \right)\rho_{\alpha} - \Omega_{12}\rho_{\beta} - \Omega_{s} \right\},\\
\nonumber\\
\dot{\rho}_{\beta} &= -i \left\{\left( \Delta_{s} - i\gamma_{2} \right)\rho_{\beta} - \Omega_{12}\rho_{\alpha} \right\},
\end{align}
\end{subequations}
with $\Delta_{s} = \omega - \omega_{s}$. Note the equations above are completely equivalent to eqs.\eqref{EvolveEITB} for $\rho_{31}$ and $\rho_{21}$, respectively, if we consider the stationary solution $\dot{\rho}_{\alpha,\beta}(t) = 0$. It shows that the dynamics of both systems, EIT and coupled oscillators, are also equivalent with $\rho_{31} \equiv \rho_{\alpha}$ and $\rho_{21} \equiv \rho_{\beta}$. In the steady state eqs.\eqref{HRhosB} gives for $\rho_{\alpha}$,
\begin{align}\label{HamRho}
\rho_{\alpha}(\omega_{s}) =\frac{\Omega_{s}\left(\Delta_{s} - i\gamma_{2} \right)}{\left(\Delta_{s} - i\gamma_{1}\right)\left(\Delta_{s} - i\gamma_{2}\right) - \Omega_{12}^{2}},
\end{align}
showing that $\rho_{\alpha} = \rho_{co}$ for $\Delta_{1,2} = \Delta_{s}$ in eq.\eqref{e8}, as expected, once the Hamiltonian is equivalent to the Newtonian formalism.

\subsection {The classical dark state}

Here we explain the Physics underlying the classical dark state for two coupled harmonic oscillators. For this we used the concepts of normal coordinates and normal modes to describe the collective motion of the system. This state is obtained when oscillator 1 is driven resonantly ($\omega_{s} = \omega_{1}$) by the harmonic force $F_{s}(t)$, causing the cancelation of the reduced mechanical susceptibility $\tilde{\chi}_{M}(\omega_{s})=\rho_{co}(\omega_{s})$ defined in Sec.\ref{SecEIT}. We consider the simple case where $m_{1,2} = m$ and $\omega_{1,2} = \omega$.

From the definition of the normal coordinates
\begin{subequations}\label{MNXMm}
\begin{eqnarray}
X_{+} &=&\left(x_{1}+x_{2}\right) /\sqrt{2}, \\
X_{-} &=&\left(x_{1}-x_{2}\right) /\sqrt{2},
\end{eqnarray}
\end{subequations}
and the normal momenta
\begin{subequations}\label{MNPMm}
\begin{eqnarray}
P_{+} &=&\left(p_{1}+p_{2}\right) /\sqrt{2}, \\
P_{-} &=&\left(p_{1}-p_{2}\right) /\sqrt{2},
\end{eqnarray}
\end{subequations}
the coupled Hamiltonian given in eq.\eqref{HamiltA}, Appendix A, is now written as a combination of two uncoupled forced harmonic oscillators:
\begin{eqnarray}
H_{nm} = \sum_{i=+,-} \left( \frac{P_{i}^{2}}{2m}+\frac{1}{2}m\omega_{i}^{2}X_{i}^{2} - \frac{\sqrt{2}}{2}F_{s}(t)X_{i} \right), 
\end{eqnarray}
where $\omega_{+} = \sqrt{k/m}$ and $\omega_{-} = \sqrt{\omega^{2}_{+} + 2\omega^{2}_{12}}$ are the resonance frequencies of the two normal modes of the system. Those are usually labeled as symmetric ($NM_{(+)}$) and asymmetric ($NM_{(-)}$) modes, owing the collective motion performed by each other. In $NM_{(+)}$ both masses move in phase with frequency $\omega_{+}$ and the amplitudes are equal. In $NM_{(-)}$ both masses move oppositely, outward and then inward, with frequency $\omega_{-}$, which is higher than $\omega_{+}$ because the middle spring is now stretched or compressed adding its effect to the restoring force.

As we have seen, the equations of motion \eqref{classicA} described in Sec.\ref{SecEIT} are obtained by adding the damping force $-\eta_{j}\dot{x}_{j}$ to the resultant force of each oscillator, with $\eta_{j} = 2m\gamma_{j}$ $(j = 1, 2)$. From eqs.\eqref{HamiltA}, \eqref{MNXMm}, \eqref{MNPMm} and the Hamilton equation, 
\begin{eqnarray}
\dot{p}_{j} = -\frac{\partial H}{\partial x_{j}} - 2m\gamma_{j}\dot{x}_{j},
\end{eqnarray}
the equations of motion for the normal coordinates are
\begin{subequations}\label{NCX}
\begin{eqnarray}
\ddot{X}_{+} + \Gamma\dot{X}_{+} + \gamma\dot{X}_{-} + \omega_{+}^{2}X_{+} &=& \frac{F_{s}(t)}{m\sqrt{2}}, \\
\ddot{X}_{-} + \gamma\dot{X}_{+} + \Gamma\dot{X}_{-} + \omega_{-}^{2}X_{-} &=& \frac{F_{s}(t)}{m\sqrt{2}},
\end{eqnarray}
\end{subequations}
with $\Gamma =\left(\gamma_{1} + \gamma_{2}\right)$ and $\gamma =\left(\gamma_{1} - \gamma_{2}\right)$. Note that the collective motions, provided by the normal modes, become uncoupled for $\gamma_{1} = \gamma_{2}$, once the coupling is performed through the asymmetric dissipation $\gamma$.

As before, we assume that the steady-state solution for the normal coordinates has the form $X_{i} = N_{i}e^{-i\omega_{s}t} + c.c.$, which conducts to the relationship
\begin{eqnarray}\label{relNM}
N_{+} = \left[\frac{\omega_{-}^{2} - \omega_{s}^{2} + 2i\gamma_{2}\omega_{s}}{\omega_{+}^{2} - \omega_{s}^{2} + 2i\gamma_{2}\omega_{s}}\right]N_{-}.
\end{eqnarray}

Using the explicit values of $\omega_{+}$ and $\omega_{-}$ defined previously, the classical dark state is obtained when $\omega_{s} = \omega$, with $\omega^2 = \omega_{+}^2 + \omega_{12}^2$. Then,
\begin{eqnarray}\label{Nmodes}
N_{+} = \left[\frac{\omega_{12}^{2} - 2i\gamma_{2}\omega}{-\omega_{12}^{2} - 2i\gamma_{2}\omega}\right]N_{-}.
\end{eqnarray}

Note that the system is pumped in a region of high interference between the normal modes, once $\omega_{s} = \omega$ is a frequency in the range between $\omega_{+}$ and $\omega_{-}$. To see how this state looks like we have to apply the classical analog for the EIT condition, which is $\Omega_{12}>>\Omega_{s}$ and $\gamma_{2}<<\gamma_{1}$, see Sec.\ref{SecEIT} for more details. For $\gamma_{2} \rightarrow 0$, eq.\eqref{Nmodes} provides $N_{+} = - N_{-}$ and consequently $X_{+} = - X_{-}$. From eqs.\eqref{MNXMm} it can be shown readily that $x_{1} = \sqrt{2}/2 \left(X_{+}+X_{-}\right)$ and $x_{2} = \sqrt{2}/2 \left(X_{+}-X_{-}\right)$. Note that the displacement of both oscillators can be described as a superposition of the two normal modes of the system. In this particular case we have $x_{1} = 0$ and $x_{2} \neq 0$. Then, the classical dark state is obtained when oscillator 1 stays stationary while oscillator 2 oscillates harmonically, meaning that it is featured by zero absorption power of oscillator 1. From eq.\eqref{e8} wee see that $\rho_{co}(\omega_{s}) \propto \left(N_{+}+N_{-}\right)$, justifying why $\rho_{co}(\omega_{s}) = 0$ throughout the paper for zero detuning, like in Fig.\ref{FigCOEIT}.

The first EIT-like condition $\Omega_{s}<<\Omega_{12}$ is demonstrated for $\gamma_{2} \neq 0$. If $\gamma_{2}<<1$, eq.\eqref{Nmodes} becomes
\begin{eqnarray}\label{NmodesB}
N_{+} = - \left[1 - \frac{4i\gamma_{2}\omega}{\omega_{12}^{2}}\right]N_{-}.
\end{eqnarray}
The condition above is equivalent to $\gamma_{2}<<\gamma_{1}$, because all parameters of the system are scaled to $\gamma_{1}$. In this case the classical dark state remains observable when $k_{12}>>k_{1}$, which implies that $\omega \approx \omega_{12} = \sqrt{k_{12}/m}$ and then $N_{+} \approx -N_{-}$. If the frequency $\omega$ of the driven oscillator is taken from the expressions for the classical pumping $\Omega_{s} = \sqrt{F^{2}/2m\omega}$ and coupling $\Omega_{12} = \omega_{12}^{2}/2\omega$ rates, we have $\Omega_{s} = F\sqrt{\Omega_{12}/k_{12}}$. In the usual approximation of small oscillations the strength of the force, given by the amplitude $F$, is very small. Then, if $k_{12}>>1$, which is fulfilled for $k_{12}>>k_{1}$, the condition $\Omega_{s}<<\Omega_{12}$ must be prescribed for $\gamma_{2} \neq 0$, in analogy to the EIT system, where $\Omega_{p}<<\Omega_{c}$ since $\gamma_{2}<<\gamma_{31}$ for nonvanishing $\gamma_{2}$.

Thus, we show that the classical dark state is caused by a destructive interference between the normal modes $NM_{(\pm)}$ in the displacement of oscillator 1, and it is observed in analogous conditions with the dark state of the EIT system. The normal modes description performed here can be extended to the case of three coupled harmonic oscillators, as discussed in Sec.III, where the classical dark state is defined according to the configuration of the system.


\begin{thebibliography}{99}
\bibitem {Multilevel}Although the complexity of realistic atoms, which are composed by a multilevel energy structure, optical pumping techniques are used to adequately produce two, three or four-level atoms. This makes the modeling of simple energy structures configurations, such as the systems presented here, reasonable.
\end{thebibliography}
\end{document}